\documentclass[11pt]{article}
\usepackage{amssymb}

\usepackage[totalwidth=460truept,totalheight=600truept]{geometry}
\usepackage{latexsym,amssymb,amsmath,graphicx,accents,eucal,slashed,subfigure,stmaryrd}
\usepackage{dsfont}
\usepackage[T1]{fontenc}
\usepackage{hyperref}

\def\theequation{\arabic{section}.\arabic{equation}}

\renewcommand{\theequation}{\thesection.\arabic{equation}}
\global\arraycolsep=1truept

\numberwithin{equation}{section}
\renewcommand{\theequation}{\arabic{section}.\arabic{equation}}

\newcommand{\ul}[1]{\mkern2mu\underline{\mkern-2mu\smash{#1}\mkern-2mu}\mkern2mu }
\newcommand{\hatbar}[1]{\hat{\bar{#1}}}
\newcommand{\brevebar}[1]{\breve{\bar{#1}}}
\newtheorem{theorem}{Theorem}
\newtheorem{corollary}[theorem]{Corollary}

\begin{document}

\bigskip \hfill IFUP-TH 2013/21

\phantom{C} \vskip 1.4truecm

\begin{center}
{\huge \textbf{Background Field Method,}}

\vskip .5truecm

{\huge \textbf{Batalin-Vilkovisky Formalism And}}

\vskip .5truecm

{\huge \textbf{Parametric Completeness Of\ Renormalization}}

\vskip 1truecm

\textsl{Damiano Anselmi} \vskip .2truecm

\textit{Dipartimento di Fisica ``Enrico Fermi'', Universit\`{a} di Pisa, }

\textit{and INFN, Sezione di Pisa,}

\textit{Largo B. Pontecorvo 3, I-56127 Pisa, Italy,}

\vskip .2truecm

damiano.anselmi@df.unipi.it

\vskip 1.5truecm

\textbf{Abstract}
\end{center}

\medskip

We investigate the background field method with the
Batalin-Vilkovisky formalism, to generalize known results, study the parametric completeness of general gauge theories and
achieve a better understanding of several properties. In particular, we
study renormalization and gauge dependence to all orders. Switching between
the background field approach and the usual approach by means of canonical
transformations, we prove parametric completeness without making use of
cohomological theorems; namely we show that if the starting classical action is
sufficiently general all divergences can be subtracted by means of parameter
redefinitions and canonical transformations. Our approach applies to
renormalizable and nonrenormalizable theories that are manifestly free of
gauge anomalies and satisfy the following assumptions: the gauge algebra is
irreducible and closes off shell, the gauge transformations are linear
functions of the fields, and closure is field independent. Yang-Mills
theories and quantum gravity in arbitrary dimensions are included, as well
as effective and higher-derivative versions of them, but several other
theories, such as supergravity, are left out.

\vskip 1truecm

\vfill\eject

\section{Introduction}

\setcounter{equation}{0}

The background field method \cite{dewitt,abbott} is a convenient tool to
quantize gauge theories and make explicit calculations, particularly when it
is used in combination with the dimensional-regularization technique. It
amounts to choosing a nonstandard gauge fixing in the conventional approach
and, among its virtues, it keeps the gauge transformations intact under
renormalization. However, it takes advantage of properties that only
particular classes of theories have.

The Batalin-Vilkovisky formalism \cite{bata} is also useful for quantizing
general gauge theories, especially because it collects all ingredients of
infinitesimal gauge symmetries in a single identity, the master equation,
which remains intact through renormalization, at least in the absence of gauge
anomalies.

Merging the background field method with the Batalin-Vilkovisky formalism is
not only an interesting theoretical subject \textit{per se}, but can also
offer a better understanding of known results, make us appreciate aspects that have
been overlooked, generalize the validity of crucial theorems about the
quantization of gauge theories and renormalization, and help us address open
problems. For example, an important issue concerns the generality of the
background field method. It would be nice to formulate a unique treatment
for all gauge theories, renormalizable and nonrenormalizable, unitary and
higher derivative, with irreducible or reducible gauge algebras that close
off shell or only on shell. However, we will see that at this stage it is
not possible to achieve that goal, due to some intrinsic features of the
background field method.

Another important issue that we want to emphasize more than has been done
so far is the problem of \textit{parametric completeness} in general gauge theories \cite{regnocoho}.
To ensure renormalization-group (RG) invariance, all divergences must be
subtracted by redefining parameters and making canonical transformations. When a theory contains all independent parameters necessary to achieve
this goal, we say that it is parametrically complete. The
RG-invariant renormalization of divergences may require the introduction of
missing Lagrangian terms, multiplied by new physical constants, or even
deform the symmetry algebra in nontrivial ways. However, in
nonrenormalizable theories such as quantum gravity and supergravity it is not obvious
that the action can indeed be adjusted to achieve parametric completeness.
One way to deal with this problem is to classify the whole cohomology of
invariants and hope that the solution satisfies suitable properties. This method requires
lengthy technical proofs that must be done case by case \cite{coho}, and
therefore lacks generality. Another way is to let renormalization build the
new invariants automatically, as shown in ref. \cite{regnocoho}, with an
algorithm that is able to iteratively extend the classical action converting
divergences into finite counterterms. However, that procedure is mainly a
theoretical tool, because although very general and conceptually minimal, it
is practically unaffordable. Among the other things, it leaves the
possibility that renormalization may dynamically deform the gauge symmetry
in physically observable ways. A third possibility is the one we are going
to treat here, taking advantage of the background field method. Where it
applies, it makes cohomological classifications unnecessary and excludes
that renormalization may dynamically deform the symmetry in observable ways.
Because of the intrinsic properties of the background field method, the
approach of this paper, although general enough, is not exhaustive. It is
general enough because it includes the gauge symmetries we need for physical
applications, namely Abelian and non-Abelian Yang-Mills symmetries, local
Lorentz symmetry and invariance under general changes of coordinates. At the
same time, it is not exhaustive because it excludes other potentially
interesting symmetries, such as local supersymmetry.

To be precise, our results hold for every gauge symmetry that satisfies the
following properties: the algebra of gauge transformations 
($i$) closes off shell and ($ii$) is
irreducible; moreover ($iii$) there exists a choice of field variables where
the gauge transformations $\delta _{\Lambda }\phi $ of the physical fields $%
\phi $ are linear functions of $\phi $ and the closure $[\delta _{\Lambda
},\delta _{\Sigma }]=\delta _{[\Lambda ,\Sigma ]}$ of the algebra is $\phi $
 independent. We expect that with some technical work it will be possible to
extend our results to theories that do not satisfy assumption ($ii$), but our
impression is that removing assumptions ($i$) and ($iii$)\ will be much
harder, if not impossible.

In this paper we also assume that the theory is manifestly free of gauge
anomalies. Our results apply to renormalizable and nonrenormalizable
theories that satisfy the assumptions listed so far, among which are QED,
Yang-Mills theories, quantum gravity and Lorentz-violating gauge theories \cite
{lvgauge}, as well as effective \cite{weinberg}, higher-derivative \cite
{stelle} and nonlocal \cite{tombola} versions of such theories, in
arbitrary dimensions, and extensions obtained including any set of composite
fields. We recall that Stelle's proof \cite{stelle} that higher-derivative
quantum gravity is renormalizable was incomplete, because it assumed without
proof a generalization of the Kluberg-Stern--Zuber conjecture \cite{kluberg}
for the cohomological problem satisfied by counterterms. Even the
cohomological analysis of refs. \cite{coho} does not directly apply to
higher-derivative quantum gravity, because the field equations of
higher-derivative theories are not equal to perturbative corrections of the
ordinary field equations. These remarks show that our results are quite
powerful, because they overcome a number of difficulties that otherwise need
to be addressed case by case.

Strictly speaking, our results, in their present form, do not apply to
chiral theories, such as the Standard Model coupled to quantum gravity,
where the cancellation of anomalies is not manifest. Nevertheless, since all
other assumptions we have made concern just the forms of gauge symmetries,
not the forms of classical actions, nor the limits around which perturbative
expansions are defined, we expect that our results can be extended to all
theories involving the Standard Model or Lorentz-violating extensions of it 
\cite{kostelecky,LVSM}. However, to make derivations more easily
understandable it is customary to first make proofs in the framework where
gauge anomalies are manifestly absent, and later extend the results by means
of the Adler-Bardeen theorem \cite{adlerbardeen}. We follow the tradition on
this, and plan to devote a separate investigation to anomaly cancellation.

Although some of our results are better understandings or generalizations of
known properties, we do include them for the sake of clarity and
self-consistence. We think that our formalism offers insight on the issues
mentioned above and gives a more satisfactory picture. In particular, the
fact that background field method makes cohomological classifications
unnecessary is something that apparently has not been appreciated enough so
far. Moreover, our approach points out the limits of applicability of the
background field method.

To achieve parametric completeness we proceed in four basic steps. First, we
study renormalization to all orders subtracting divergences ``as they
come'', which means without worrying whether the theory contains enough
independent parameters for RG invariance or not. Second, we study how the
renormalized action and the renormalized $\Gamma $ functional depend on the
gauge fixing, and work out how the renormalization algorithm maps a
canonical transformation of the classical theory into a canonical
transformation of the renormalized theory. Third, we renormalize the
canonical transformation that continuously interpolates between the
background field approach and the conventional approach. Fourth, comparing
the two approaches we show that if
the classical action $S_{c}(\phi ,\lambda )$ contains all gauge invariant
terms determined by the starting gauge symmetry, then there exists a canonical transformation $%
\Phi ,K\rightarrow \hat{\Phi},\hat{K}$ such that 
\begin{equation}
S_{R\hspace{0.01in}\text{min}}(\Phi ,\ul{\phi },K)=S_{c}(\hat{\phi}+%
\ul{\phi },\tau (\lambda ))-\int R^{\alpha }(\hat{\phi}+\ul{%
\phi },\hat{C})\hat{K}_{\alpha },  \label{key0}
\end{equation}
where $S_{R\hspace{0.01in}\text{min}}$ is the renormalized action with the
gauge-fixing sector switched off, $\Phi ^{\alpha }=\{\phi ,C\}$ are the
fields ($C$ being the ghosts), $K_{\alpha }$ are the sources for the $\Phi^{\alpha}$
transformations $R^{\alpha }(\Phi )$, $\ul{\phi }$ are the background
fields, $\lambda $ are the
physical couplings and $\tau (\lambda )$ are $\lambda $ redefinitions. Identity (\ref{key0}) shows
that all divergences can be renormalized by means of parameter redefinitions
and canonical transformations, which proves parametric completeness. Power
counting may or may not restrict the form of $S_{c}(\phi ,\lambda )$.

Basically, under the assumptions we have made the background transformations
do not renormalize, and the quantum fields $\phi $ can be switched off and
then restored from their background partners $\ul{\phi }$.
Nevertheless, the restoration works only up to a canonical transformation,
which gives (\ref{key0}). The story is a bit more complicated than this, but
this simplified version is enough to appreciate the main point. However,
when the assumptions we have made do not hold, the argument fails, which
shows how peculiar the background field method is. Besides giving explicit
examples where the construction works, we address some problems that arise
when the assumptions listed above are not satisfied.

A somewhat different approach to the background field method in the
framework of the Batalin-Vilkovisky formalism exists in the literature. In
refs. \cite{quadri} Binosi and Quadri considered the most general variation $%
\delta \ul{A}=\Omega $ of the background gauge field $\ul{A}$
in Yang-Mills theory, and obtained a modified Batalin-Vilkovisky master
equation that controls how the functional $\Gamma $ depends on $\ul{A}
$. Instead, here we introduce background copies of both physical fields and
ghosts, which allows us to split the symmetry transformations into ``quantum
transformations'' and ``background transformations''. The master equation is
split into the three identities (\ref{treide}), which control invariances
under the two types of transformations.

The paper is organized as follows. In section 2 we formulate our approach
and derive its basic properties, emphasizing the assumptions we make and why
they are necessary. In section 3 we renormalize divergences to all orders,
subtracting them ``as they come''. In section 4 we derive the basic
differential equations of gauge dependence and integrate them, which allows
us to show how a renormalized canonical transformation emerges from its
tree-level limit. In section 5 we derive (\ref{key0}) and prove parametric
completeness. In section 6 we give two examples, non-Abelian Yang-Mills
theory and quantum gravity. In section 7 we make remarks about parametric
completeness and recapitulate where we stand now on this issue. Section 8
contains our conclusions, while the appendix collects several theorems and
identities that are used in the paper.

We use the dimensional-regularization technique and the minimal subtraction scheme. Recall that the functional
integration measure is invariant with respect to perturbatively local
changes of field variables. Averages $\langle \cdots \rangle $ always denote
the sums of \textit{connected} Feynman diagrams. We use the Euclidean
notation in theoretical derivations and switch to Minkowski spacetime in the
examples.

\section{Background field method and Batalin-Vilkovisky formalism}

\setcounter{equation}{0}

In this section we formulate our approach to the background field method
with the Batalin-Vilkovisky formalism. To better appreciate the arguments
given below it may be useful to jump back and forth between this section and
section 6, where explicit examples are given.

If the gauge algebra closes off shell, there exists a canonical
transformation that makes the solution $S(\Phi ,K)$ of the master equation $%
(S,S)=0$ depend linearly on the sources $K$. We write 
\begin{equation}
S(\Phi ,K)=\mathcal{S}(\Phi )-\int R^{\alpha }(\Phi )K_{\alpha }.
\label{solp}
\end{equation}
The fields $%
\Phi ^{\alpha }=\{\phi ^{i},C^{I},\bar{C}^{I},B^{I}\}$ are made of
physical fields $\phi ^{i}$, ghosts $C^{I}$ (possibly including
ghosts of ghosts and so on), antighosts $\bar{C}^{I}$ and Lagrange
multipliers $B^{I}$ for the gauge fixing. Moreover, $K_{\alpha }=\{K_{\phi
}^{i},K_{C}^{I},K_{\bar{C}}^{I},K_{B}^{I}\}$ are the sources associated with
the symmetry transformations $R^{\alpha}(\Phi)$ of the fields $\Phi ^{\alpha }$, while 
\[
\mathcal{S}(\Phi )=S_{c}(\phi )+(S,\Psi ) 
\]
is the sum of the classical action $S_{c}(\phi )$ plus the gauge fixing,
which is expressed as the antiparenthesis of $S$ with a $K$-independent
gauge fermion $\Psi (\Phi )$. We recall that the antiparentheses are defined
as 
\[
(X,Y)=\int \left\{ \frac{\delta _{r}X}{\delta \Phi ^{\alpha }}\frac{\delta
_{l}Y}{\delta K_{\alpha }}-\frac{\delta _{r}X}{\delta K_{\alpha }}\frac{%
\delta _{l}Y}{\delta \Phi ^{\alpha }}\right\} ,
\]
where the summation over the index $\alpha $ is understood. The integral is over spacetime points associated with repeated indices. 

The non-gauge-fixed action 
\begin{equation}
S_{\text{min}}(\Phi ,K)=S_{c}(\phi )-\int R_{\phi }^{i}(\phi ,C)K_{\phi
}^{i}-\int R_{C}^{I}(\phi ,C)K_{C}^{I},  \label{smin}
\end{equation}
obtained by dropping antighosts, Lagrange multipliers and their sources, also
solves the master equation, and is called the minimal solution. Antighosts $\bar{%
C}$ and Lagrange multipliers $B$ form trivial gauge systems, and typically
enter (\ref{solp}) by means of the gauge fixing $(S,\Psi )$ and a
contribution 
\begin{equation}
\Delta S_{\text{nm}}=-\int B^{I}K_{\bar{C}}^{I},  \label{esto}
\end{equation}
to $-\int R^{\alpha }K_{\alpha }$.

Let $\mathcal{R}^{\alpha }(\Phi ,C)$ denote the transformations the fields $%
\Phi ^{\alpha }$ would have if they were matter fields. Each function $%
\mathcal{R}^{\alpha }(\Phi ,C)$ is a bilinear form of $\Phi ^{\alpha }$ and $%
C$. Sometimes, to be more explicit, we also use the notation $\mathcal{R}_{%
\bar{C}}^{I}(\bar{C},C)$ and $\mathcal{R}_{B}^{I}(B,C)$ for $\bar{C}$ and $B$%
, respectively. It is often convenient to replace (\ref{esto}) with the
alternative nonminimal extension 
\begin{equation}
\Delta S_{\text{nm}}^{\prime }=-\int \left( B^{I}+\mathcal{R}_{\bar{C}}^{I}(%
\bar{C},C)\right) K_{\bar{C}}^{I}-\int \mathcal{R}_{B}^{I}(B,C)K_{B}^{I}.
\label{estobar}
\end{equation}
For example, in Yang-Mills theories we have 
\[
\Delta S_{\text{nm}}^{\prime }=-\int \left( B^{a}-gf^{abc}C^{b}\bar{C}%
^{c}\right) K_{\bar{C}}^{a}+g\int f^{abc}C^{b}B^{c}K_{B}^{a} 
\]
and in quantum gravity 
\begin{equation}
\Delta S_{\text{nm}}^{\prime }=-\int \left( B_{\mu }+\bar{C}_{\rho }\partial
_{\mu }C^{\rho }-C^{\rho }\partial _{\rho }\bar{C}_{\mu }\right) K_{\bar{C}%
}^{\mu }+\int \left( B_{\rho }\partial _{\mu }C^{\rho }+C^{\rho }\partial
_{\rho }B_{\mu }\right) K_{B}^{\mu },  \label{nmqg}
\end{equation}
where $C^{\mu }$ are the ghosts of diffeomorphisms.

Observe that (\ref{estobar}) can be obtained from (\ref{esto}) making the
canonical transformation generated by 
\[
F_{\text{nm}}(\Phi ,K^{\prime })=\int \Phi ^{\alpha }K_{\alpha }^{\prime
}+\int \mathcal{R}_{\bar{C}}^{I}(\bar{C},C)K_{B}^{I\hspace{0.01in}\prime }. 
\]
Requiring that $F_{\text{nm}}$ indeed give (\ref{estobar}) we get the
identities 
\begin{equation}
\mathcal{R}_{B}^{I}(B,C)=-\int B^{J}\frac{\delta _{l}}{\delta \bar{C}^{J}}%
\mathcal{R}_{\bar{C}}^{I}(\bar{C},C),\qquad \int \left( R_{C}^{J}\frac{%
\delta _{l}}{\delta C^{J}}+\mathcal{R}_{\bar{C}}^{J}(\bar{C},C)\frac{\delta
_{l}}{\delta \bar{C}^{J}}\right) \mathcal{R}_{\bar{C}}^{I}(\bar{C},C)=0,
\label{iddo}
\end{equation}
which can be easily checked both for Yang-Mills theories and gravity. In
this paper the notation $R^{\alpha }(\Phi )$ refers to the
field transformations of (\ref{smin}) plus those of the nonminimal
extension (\ref{esto}), while $\bar{R}^{\alpha }(\Phi )$ refers to the
transformations of (\ref{smin}) plus (\ref{estobar}).


\subsection{Background field action}

To apply the background field method, we start from the gauge invariance of
the classical action $S_{c}(\phi )$, 
\begin{equation}
\int R_{c}^{i}(\phi ,\Lambda )\frac{\delta _{l}S_{c}(\phi )}{\delta \phi ^{i}%
}=0,  \label{lif}
\end{equation}
where $\Lambda $ are the arbitrary functions that parametrize the gauge
transformations $\delta \phi ^{i}=R_{c}^{i}$. Shifting the fields $\phi $ by
background fields $\ul{\phi }$, and introducing arbitrary background
functions $\ul{\Lambda }$ we can write the identity 
\[
\int \left[ R_{c}^{i}(\phi +\ul{\phi },\Lambda )+X^{i}\right] \frac{%
\delta _{l}S_{c}(\phi +\ul{\phi })}{\delta \phi ^{i}}+\int \left[
R_{c}^{i}(\phi +\ul{\phi },\ul{\Lambda })-X^{i}\right] \frac{%
\delta _{l}S_{c}(\phi +\ul{\phi })}{\delta \ul{\phi }^{i}}=0, 
\]
which is true for arbitrary functions $X^{i}$. If we choose 
\[
X^{i}=R_{c}^{i}(\phi +\ul{\phi },\ul{\Lambda })-R_{c}^{i}(%
\ul{\phi },\ul{\Lambda }), 
\]
the transformations of the background fields contain only background fields
and coincide with $R_{c}^{i}(\ul{\phi },\ul{\Lambda })$. We
find 
\begin{equation}
\int \left[ R_{c}^{i}(\phi +\ul{\phi },\Lambda +\ul{\Lambda }%
)-R_{c}^{i}(\ul{\phi },\ul{\Lambda })\right] \frac{\delta
_{l}S_{c}(\phi +\ul{\phi })}{\delta \phi ^{i}}+\int R_{c}^{i}(%
\ul{\phi },\ul{\Lambda })\frac{\delta _{l}S_{c}(\phi +%
\ul{\phi })}{\delta \ul{\phi }^{i}}=0.  \label{bas}
\end{equation}

Thus, denoting background quantities by means of an underlining, we are led
to consider the action 
\begin{equation}
S(\Phi ,\ul{\Phi },K,\ul{K})=S_{c}(\phi +\ul{\phi }%
)-\int R^{\alpha }(\Phi +\ul{\Phi })K_{\alpha }-\int R^{\alpha }(%
\ul{\Phi })(\ul{K}_{\alpha }-K_{\alpha }),  \label{sback}
\end{equation}
which solves the master equation $\llbracket S,S\rrbracket =0$, where the
antiparentheses are defined as 
\[
\llbracket X,Y\rrbracket =\int \left\{ \frac{\delta _{r}X}{\delta \Phi
^{\alpha }}\frac{\delta _{l}Y}{\delta K_{\alpha }}+\frac{\delta _{r}X}{%
\delta \ul{\Phi }^{\alpha }}\frac{\delta _{l}Y}{\delta \ul{K}%
_{\alpha }}-\frac{\delta _{r}X}{\delta K_{\alpha }}\frac{\delta _{l}Y}{%
\delta \Phi ^{\alpha }}-\frac{\delta _{r}X}{\delta \ul{K}_{\alpha }}%
\frac{\delta _{l}Y}{\delta \ul{\Phi }^{\alpha }}\right\} . 
\]

More directly, if $S(\Phi ,K)=S_{c}(\phi )-\int R^{\alpha }(\Phi )K_{\alpha
} $ solves $(S,S)=0$, the background field can be introduced with a
canonical transformation. Start from the action 
\begin{equation}
S(\Phi ,\ul{\Phi },K,\ul{K})=S_{c}(\phi )-\int R^{\alpha
}(\Phi )K_{\alpha }-\int R^{\alpha }(\ul{\Phi })\ul{K}_{\alpha
},  \label{sback0}
\end{equation}
which obviously satisfies two master equations, one in the variables $\Phi
,K $ and the other one in the variables $\ul{\Phi },\ul{K}$. {\it A
fortiori}, it also satisfies $\llbracket S,S\rrbracket =0$. Relabeling fields
and sources with primes and making the canonical transformation generated by
the functional 
\begin{equation}
F_{\text{b}}(\Phi ,\ul{\Phi },K^{\prime },\ul{K}^{\prime
})=\int (\Phi ^{\alpha }+\ul{\Phi }^{\alpha })K_{\alpha }^{\prime
}+\int \ul{\Phi }^{\alpha }\ul{K}_{\alpha }^{\prime },
\label{casbac}
\end{equation}
we obtain (\ref{sback}), and clearly preserve $\llbracket S,S\rrbracket =0$.

The shift $\ul{\Phi }$ is called background field, while $\Phi $ is
called quantum field. We also have quantum sources $K$ and background
sources $\ul{K}$. Finally, we have background transformations, those
described by the background ghosts $\ul{C}$ or the functions $%
\ul{\Lambda }$ in (\ref{bas}), and quantum transformations, those
described by the quantum ghosts $C$ and (\ref{esto}) or the functions $%
\Lambda $ in (\ref{bas}).

The action (\ref{sback}) is not the most convenient one to study
renormalization. It is fine in the minimal sector (the one with antighosts
and Lagrange multipliers switched off), but not in the nonminimal one. Now
we describe the improvements we need to make.

\paragraph{Non-minimal sector\newline
}

So far we have introduced background copies of all fields. Nevertheless,
strictly speaking we do not need to introduce copies of the antighosts $\bar{%
C}$ and the Lagrange multipliers $B$, since we do not need to gauge-fix the
background. Thus we drop $\ul{\bar{C}}$, $\ul{B}$ and their
sources from now on, and define $\ul{\Phi }^{\alpha }=\{\ul{%
\phi }^{i},\ul{C}^{I},0,0\}$, $\ul{K}_{\alpha }=\{\ul{K}%
_{\phi }^{i},\ul{K}_{C}^{I},0,0\}$. Observe that then we have $%
R^{\alpha }(\ul{\Phi })=\bar{R}^{\alpha }(\ul{\Phi }%
)=\{R_{\phi }^{i}(\ul{\phi },\ul{C}),R_{C}^{I}(\ul{\Phi 
}),0,0\}$.

Let us compare the nonminimal sectors (\ref{esto}) and (\ref{estobar}). If
we choose (\ref{esto}), $\bar{C}$ and $B$ do not transform under background
transformations. Since (\ref{esto}) are the only terms that contain $K_{\bar{%
C}}$, they do not contribute to one-particle irreducible diagrams and do not
receive radiative corrections. Moreover, $K_{B}$ does not appear in the
action. Instead, if we choose the nonminimal sector (\ref{estobar}), namely
if we start from 
\begin{equation}
S(\Phi ,\ul{\Phi },K,\ul{K})=S_{c}(\phi )-\int \bar{R}^{\alpha
}(\Phi )K_{\alpha }-\int R^{\alpha }(\ul{\Phi })\ul{K}_{\alpha
}  \label{sback1}
\end{equation}
instead of (\ref{sback0}), the transformation (\ref{casbac}) gives the
action 
\begin{equation}
S(\Phi ,\ul{\Phi },K,\ul{K})=S_{c}(\phi +\ul{\phi }%
)-\int (\bar{R}^{\alpha }(\Phi +\ul{\Phi })-\bar{R}^{\alpha }(%
\ul{\Phi }))K_{\alpha }-\int R^{\alpha }(\ul{\Phi })\ul{%
K}_{\alpha }.  \label{sback2}
\end{equation}
In particular, using the linearity of $\mathcal{R}_{\bar{C}}^{I}$ and $%
\mathcal{R}_{B}^{I}$ in $C$, we see that (\ref{estobar}) is turned into
itself plus 
\begin{equation}
-\int \mathcal{R}_{\bar{C}}^{I}(\bar{C},\ul{C})K_{\bar{C}}^{I}-\int 
\mathcal{R}_{B}^{I}(B,\ul{C})K_{B}^{I}.  \label{bacca}
\end{equation}
Because of these new terms, $\bar{C}$ and $B$ now transform as ordinary
matter fields under background transformations. This is the correct
background transformation law we need for them. On the other hand, the
nonminimal sector (\ref{estobar}) also generates nontrivial quantum
transformations for $\bar{C}$ and $B$, which are renormalized and complicate
our derivations.

It would be better to have (\ref{estobar}) in the background sector and (\ref
{esto}) in the nonbackground sector. To achieve this goal, we make the
canonical transformation generated by 
\begin{equation}
F_{\text{nm}}^{\prime }(\Phi ,\ul{\Phi },K^{\prime },\ul{K}%
^{\prime })=\int \Phi ^{\alpha }K_{\alpha }^{\prime }+\int \ul{\Phi }%
^{\alpha }\ul{K}_{\alpha }^{\prime }+\int \mathcal{R}_{\bar{C}}^{I}(%
\bar{C},\ul{C})K_{B}^{I\hspace{0.01in}\prime }  \label{casbacca}
\end{equation}
on (\ref{sback}). Using (\ref{iddo}) again, the result is 
\begin{eqnarray}
S(\Phi ,\ul{\Phi },K,\ul{K}) &=&S_{c}(\phi +\ul{\phi }%
)-\int (R^{\alpha }(\Phi +\ul{\Phi })-R^{\alpha }(\ul{\Phi }%
))K_{\alpha }  \nonumber \\
&&-\int \mathcal{R}_{\bar{C}}^{I}(\bar{C},\ul{C})K_{\bar{C}}^{I}-\int 
\mathcal{R}_{B}^{I}(B,\ul{C})K_{B}^{I}-\int R^{\alpha }(\ul{%
\Phi })\ul{K}_{\alpha }.  \label{sbacca}
\end{eqnarray}
This is the background field action we are going to work with. It is
straightforward to check that (\ref{sbacca}) satisfies $\llbracket S,S%
\rrbracket =0$.

\paragraph{Separating the background and quantum sectors\newline
}

Now we separate the background sector from the quantum sector. To do this
properly we need to make further assumptions. 

First, we assume that there
exists a choice of field variables where the functions $R^{\alpha }(\Phi )$
are at most quadratic in $\Phi $. We call it \textit{linearity assumption}.
It is equivalent to assume that the gauge transformations $\delta _{\Lambda
}\phi ^{i}=R_{c}^{i}(\phi ,\Lambda )$ of (\ref{lif}) are linear functions of
the fields $\phi $ and closure is expressed by $\phi $-independent
identities $[\delta _{\Lambda },\delta _{\Sigma }]=\delta _{[\Lambda ,\Sigma
]}$. The linearity assumption is satisfied by all gauge symmetries of physical
interest, such as those of QED, non-Abelian Yang-Mills theory, quantum
gravity and the Standard Model. On the other hand, it is not satisfied by other
important symmetries, among which is supergravity, where the gauge
transformations either close only on shell or are not linear in the fields.

Second, we assume that the gauge algebra is irreducible, which ensures that
the set $\Phi $ contains only ghosts and not ghosts of ghosts.

Under these assumptions, we make the canonical transformation generated by 
\begin{equation}
F_{\tau }(\Phi ,\ul{\Phi },K^{\prime },\ul{K}^{\prime })=\int
\Phi ^{\alpha }K_{\alpha }^{\prime }+\int \ul{\Phi }^{\alpha }%
\ul{K}_{\alpha }^{\prime }+(\tau -1)\int \ul{C}^{I}\ul{K%
}_{C}^{I\hspace{0.01in}\prime }  \label{backghost}
\end{equation}
on the action (\ref{sbacca}). This transformation amounts to rescaling the
background ghosts $\ul{C}^{I}$ by a factor $\tau $ and their sources $%
\ul{K}_{C}^{I}$ by a factor $1/\tau $. Since we do not have
background antighosts, (\ref{backghost}) is the background-ghost-number
transformation combined with a rescaling of the background sources.

The action (\ref{sbacca}) is not invariant under (\ref{backghost}). Using
the linearity assumption it is easy to check that the transformed action $%
S_{\tau }$ is linear in $\tau $. Writing $S_{\tau }=\hat{S}+\tau \bar{S}$ we
can split the total action $S$ into the sum $\hat{S}+\bar{S}$ of a \textit{%
quantum action} $\hat{S}$ and a \textit{background action} $\bar{S}$.

Precisely, the quantum action $\hat{S}$ does not depend on the background
sources $\ul{K}$ and the background ghosts $\ul{C}$, but only
on the background copies $\ul{\phi }$ of the physical fields. We have 
\begin{equation}
\hat{S}=\hat{S}(\Phi ,\ul{\phi },K)=S_{c}(\phi +\ul{\phi }%
)-\int R^{\alpha }(\phi +\ul{\phi },C,\bar{C},B)K_{\alpha }.
\label{deco}
\end{equation}
Note that, in spite of the notation, the functions $R^{\alpha }(\Phi )$ are
actually $\bar{C}$ independent. Moreover, we find 
\begin{equation}
\bar{S}(\Phi ,\ul{\Phi },K,\ul{K})=-\int \mathcal{R}^{\alpha
}(\Phi ,\ul{C})K_{\alpha }-\int R^{\alpha }(\ul{\Phi })%
\ul{K}_{\alpha },  \label{sbar}
\end{equation}
where, for $\phi $ and $C$, 
\begin{equation}
\mathcal{R}^{\alpha }(\Phi ,\ul{C})=R^{\alpha }(\Phi +\ul{\Phi 
})-R^{\alpha }(\ul{\Phi })-R^{\alpha }(\phi +\ul{\phi },C,\bar{%
C},B).  \label{batra}
\end{equation}
These functions transform $\phi $ and $C$ as if they were matter fields and
are of course linear in $\Phi $ and $\ul{C}$. Note that formula (\ref
{batra}) does not hold for antighosts and Lagrange multipliers. In the end
all quantum fields transform as matter fields under background
transformations.

The master equation $\llbracket S,S\rrbracket =0$ decomposes into the three
identities 
\begin{equation}
\llbracket \hat{S},\hat{S}\rrbracket =\llbracket \hat{S},\bar{S}\rrbracket =%
\llbracket \bar{S},\bar{S}\rrbracket =0,  \label{treide}
\end{equation}
which we call \textit{background field master equations}.

The quantum transformations are described by $\hat{S}$ and the background
ones are described by $\bar{S}$. Background fields are inert under quantum
transformations, because $\llbracket \hat{S},\ul{\Phi }\rrbracket =0$%
. Note that 
\begin{equation}
\llbracket \hat{S},\llbracket \bar{S},X\rrbracket \rrbracket +\llbracket 
\bar{S},\llbracket \hat{S},X\rrbracket \rrbracket =0,  \label{uso}
\end{equation}
where $X$ is an arbitrary local functional. This property follows from the
Jacobi identity of the antiparentheses and $\llbracket \hat{S},\bar{S}%
\rrbracket =0$, and states that background and quantum transformations
commute.

\paragraph{Gauge-fixing\newline
}

Now we come to the gauge fixing. In the usual approach, the theory is
typically gauge-fixed by means of a canonical transformation that amounts to
replacing the action $S$ by$\ S+(S,\Psi )$, where $\Psi $ is a local
functional of ghost number $-1$ and depends only on the fields $\Phi $.
Using the background field method it is convenient to search for a $%
\ul{C}$-independent gauge-fixing functional $\Psi (\Phi ,\ul{%
\phi })$ that is also invariant under background transformations, namely
such that 
\begin{equation}
\llbracket \bar{S},\Psi \rrbracket =0.  \label{backgf}
\end{equation}
Then we fix the gauge with the usual procedure, namely we make a canonical
transformation generated by 
\begin{equation}
F_{\text{gf}}(\Phi ,\ul{\Phi },K^{\prime },\ul{K}^{\prime
})=\int \ \Phi ^{\alpha }K_{\alpha }^{\prime }+\int \ \ul{\Phi }%
^{\alpha }\ul{K}_{\alpha }^{\prime }+\Psi (\Phi ,\ul{\phi }).
\label{backgfgen}
\end{equation}
Because of (\ref{backgf}) the gauge-fixed action reads 
\begin{equation}
S_{\text{gf}}=\hat{S}+\bar{S}+\llbracket \hat{S},\Psi \rrbracket .
\label{fgback}
\end{equation}
Defining $\hat{S}_{\text{gf}}=\hat{S}+\llbracket \hat{S},\Psi \rrbracket $,
identities (\ref{treide}), (\ref{uso}) and (\ref{backgf}) give $\llbracket 
\hat{S}_{\text{gf}},\hat{S}_{\text{gf}}\rrbracket =\llbracket 
\hat{S}_{\text{gf}},\bar{S}\rrbracket =0$, so it is just like gauge-fixing $%
\hat{S}$. Since both $\hat{S}$ and $\Psi $ are $\ul{K}$ and $%
\ul{C}$ independent, $\hat{S}_{\text{gf}}$ is also $\ul{K}$
and $\ul{C}$ independent. Observe that the canonical transformations (%
\ref{backghost}) and (\ref{backgfgen}) commute; therefore we can safely
apply the transformation (\ref{backghost}) to the gauge-fixed action. A
gauge fixing satisfying (\ref{backgf}) is called \textit{%
background-preserving gauge fixing}.

In some derivations of this paper the background field master equations (\ref
{treide}) are violated in intermediate steps; therefore we need to prove
properties that hold more generally. Specifically, consider an action 
\begin{equation}
S(\Phi ,\ul{\Phi },K,\ul{K})=\hat{S}(\Phi ,\ul{\phi }%
,K)+\bar{S}(\Phi ,\ul{\Phi },K,\ul{K}),  \label{assu}
\end{equation}
equal to the sum of a $\ul{K}$- and $\ul{C}$-independent
``quantum action'' $\hat{S}$, plus a ``background action'' $\bar{S}$ that
satisfies the following requirements: ($i$)\ it is a linear function of the
quantum fields $\Phi $, ($ii$) it gets multiplied by $\tau $ when applying the
canonical transformation (\ref{backghost}), and ($iii$) $\delta _{l}\bar{S}%
/\delta \ul{K}_{\alpha }$ is $\Phi $ independent. In particular,
requirement ($ii$) implies that $\bar{S}$ vanishes at $\ul{C}=0$.
Since $\bar{S}$ is a linear function of $\Phi $, it does not contribute to
one-particle irreducible diagrams. Since $\hat{S}$ does not depend on $%
\ul{C}$, while $\bar{S}$ vanishes at $\ul{C}=0$, $\bar{S}$
receives no radiative corrections. Thus the $\Gamma$ functional associated with the action (\ref{assu}) satisfies
\begin{equation}
\Gamma (\Phi ,\ul{\Phi },K,\ul{K})=\hat{\Gamma}(\Phi ,%
\ul{\phi },K)+\bar{S}(\Phi ,\ul{\Phi },K,\ul{K}).
\label{becco}
\end{equation}
Moreover, thanks to theorem \ref{thb} of the appendix we have the general
identity 
\begin{equation}
\llbracket \Gamma ,\Gamma \rrbracket =\langle \llbracket S,S\rrbracket %
\rangle ,  \label{univ}
\end{equation}
under the sole assumption that $\delta _{l}S/\delta \ul{K}_{\alpha }$
is $\Phi $ independent.

Applying the canonical transformation (\ref{backghost}) to $\Gamma $ we find 
$\Gamma _{\tau }=\hat{\Gamma}+\tau \bar{S}$, so (\ref{univ}) gives the
identities 
\begin{equation}
\llbracket \hat{\Gamma},\hat{\Gamma}\rrbracket =\langle \llbracket \hat{S},%
\hat{S}\rrbracket \rangle ,\qquad \llbracket \bar{S},\hat{\Gamma}\rrbracket %
=\langle \llbracket \bar{S},\hat{S}\rrbracket \rangle .  \label{give}
\end{equation}
When $\llbracket S,S\rrbracket =0$ we have 
\begin{equation}
\llbracket \Gamma ,\Gamma \rrbracket =\llbracket \hat{\Gamma},\hat{\Gamma}%
\rrbracket =\llbracket \bar{S},\hat{\Gamma}\rrbracket =0.  \label{msb}
\end{equation}

Observe that, thanks to the linearity assumption, an $\bar{S}$ equal to (\ref
{sbar}) satisfies the requirements of formula (\ref{assu}).

Now we give details about the background-preserving gauge fixing we pick for the action (\ref{sbacca}). It is convenient to choose gauge-fixing functions $G^{Ii}(\ul{\phi }%
,\partial )\phi ^{i}$ that are linear in the quantum fields $\phi $, where $%
G^{Ii}(\ul{\phi },\partial )$ may contain derivative operators.
Precisely, we choose the gauge fermion 
\begin{equation}
\Psi (\Phi ,\ul{\phi })=\int \bar{C}^{I}G^{Ii}(\ul{\phi }%
,\partial )\phi ^{i},  \label{psiback}
\end{equation}
and assume that it satisfies (\ref{backgf}). A more common choice would be (see (\ref{seeym}) for Yang-Mills theory) 
\[
\Psi (\Phi ,\ul{\phi })=\int \bar{C}^{I}\left( G^{Ii}(\ul{\phi 
},\partial )\phi ^{i}+\xi _{IJ}B^{J}\right) , 
\]
where $\xi _{IJ}$ are gauge-fixing parameters. In this case, when we
integrate the $B$ fields out the expressions $G^{Ii}(\ul{\phi }%
,\partial )\phi ^{i}$ get squared. However, (\ref{psiback}) is better for
our purposes, because it makes the canonical transformations (\ref{casbacca}%
) and (\ref{backgfgen}) commute with each other. We call the choice (\ref
{psiback}) \textit{regular Landau gauge}. The gauge-field propagators
coincide with the ones of the Landau gauge. Nevertheless, while the usual
Landau gauge (with no $B$'s around) is singular, here gauge fields are part
of multiplets that include the $B$'s, therefore (\ref{psiback}) is regular.

In the regular Landau gauge, using (\ref{backgf}) and applying (\ref
{backgfgen}) to (\ref{sbacca}) we find 
\begin{equation}
S_{\text{gf}}=\hat{S}_{\text{gf}}+\bar{S}=S_{c}(\phi +\ul{\phi }%
)-\int R^{\alpha }(\phi +\ul{\phi },C,\bar{C},B)\tilde{K}_{\alpha
}-\int \mathcal{R}^{\alpha }(\Phi ,\ul{C})K_{\alpha }-\int R^{\alpha
}(\ul{\Phi })\ul{K}_{\alpha },  \label{sbaccagf}
\end{equation}
where the tilde sources $\tilde{K}_{\alpha }$ coincide with $K_{\alpha }$
apart from $\tilde{K}_{\phi }^{i}$ and $\tilde{K}_{\bar{C}}^{I}$, which are 
\begin{equation}
\tilde{K}_{\phi }^{i}=K_{\phi }^{i}-\bar{C}^{I}G^{Ii}(\ul{\phi },-%
\overleftarrow{\partial }),\qquad \tilde{K}_{\bar{C}}^{I}=K_{\bar{C}%
}^{I}-G^{Ii}(\ul{\phi },\partial )\phi ^{i}.  \label{chif}
\end{equation}

Recalling that the functions $R^{\alpha }(\Phi )$ are $\bar{C}$ independent,
we see that $\hat{S}_{\text{gf}}$ does not depend on $K_{\phi }^{i}$ and $%
\bar{C}$ separately, but only through the combination $\tilde{K}_{\phi }^{i}$%
. Every one-particle irreducible diagram with $\bar{C}^{I}$ external legs
actually factorizes a $-\bar{C}^{I}G^{Ii}(\ul{\phi },-\overleftarrow{%
\partial })$ on those legs. Replacing one or more such objects with $K_{\phi
}^{i}$s, we obtain other contributing diagrams. Conversely, replacing one or
more $K_{\phi }^{i}$-external legs with $-\bar{C}^{I}G^{Ii}(\ul{\phi }%
,-\overleftarrow{\partial })$ we also obtain contributing diagrams.
Therefore, all radiative corrections, as well as the renormalized action $%
\hat{S}_{R}$ and the $\Gamma $ functionals $\hat{\Gamma}$ and $\hat{\Gamma}%
_{R}$ associated with the action (\ref{sbaccagf}), do not depend on $K_{\phi }^{i}$ and $\bar{C}$ separately, but only
through the combination $\tilde{K}_{\phi }^{i}$.

The only $B$-dependent terms of $\hat{S}_{\text{gf}}$, provided by $%
\llbracket S,\Psi \rrbracket $ and (\ref{esto}), are 
\begin{equation}
\Delta S_{B}\equiv -\int B^{I}\tilde{K}_{\bar{C}}^{I}=\int B^{I}\left(
G^{Ii}(\ul{\phi },\partial )\phi ^{i}-K_{\bar{C}}^{I}\right) ,
\label{chio}
\end{equation}
and are quadratic or linear in the quantum fields. For this reason, no
one-particle irreducible diagrams can contain external $B$ legs, therefore $%
\Delta S_{B}$ is nonrenormalized and goes into $\hat{S}_{R}$, $\hat{\Gamma}$
and $\hat{\Gamma}_{R}$ unmodified.

We thus learn that using linear gauge-fixing functions we can set $\bar{C}=B=0$ and
later restore the correct $\bar{C}$ and $B$ dependencies in $\hat{S}_{\text{%
gf}}$, $\hat{S}_{R}$, $\hat{\Gamma}$ and $\hat{\Gamma}_{R}$ just by replacing $%
K_{\phi }^{i}$ with $\tilde{K}_{\phi }^{i}$ and adding $\Delta S_{B}$.

From now on when no confusion can arise we drop the subscripts of $S_{\text{%
gf}}$ and $\hat{S}_{\text{gf}}$ and assume that the background field theory
is gauge-fixed in the way just explained.

\subsection{Background-preserving canonical transformations}

It is useful to characterize the most general canonical transformations $%
\Phi ,\ul{\Phi },K,\ul{K}\rightarrow \Phi ^{\prime },%
\ul{\Phi }^{\prime },K^{\prime },\ul{K}^{\prime }$ that
preserve the background field master equations (\ref{treide}) and the basic
properties of $\hat{S}$ and $\bar{S}$.

By definition, all canonical transformations preserve the antiparentheses,
so (\ref{treide}) are turned into 
\begin{equation}
\llbracket \hat{S}^{\prime },\hat{S}^{\prime }\rrbracket ^{\prime }=%
\llbracket \hat{S}^{\prime },\bar{S}^{\prime }\rrbracket ^{\prime }=%
\llbracket \bar{S}^{\prime },\bar{S}^{\prime }\rrbracket ^{\prime }=0.
\label{tretre}
\end{equation}
Moreover, $\hat{S}^{\prime }$ should be $\ul{K}^{\prime }$ and $%
\ul{C}^{\prime }$ independent, while $\bar{S}$ should be invariant,
because it encodes the background transformations. This means 
\begin{equation}
\bar{S}^{\prime }(\Phi ,\ul{\Phi },K,\ul{K})=\bar{S}(\Phi ,%
\ul{\Phi },K,\ul{K}).  \label{thesisback}
\end{equation}

We prove that a canonical transformation defined by a generating functional
of the form 
\begin{equation}
F(\Phi ,\ul{\Phi },K^{\prime },\ul{K}^{\prime })=\int \ \Phi
^{\alpha }K_{\alpha }^{\prime }+\int \ \ul{\Phi }^{\alpha }\ul{%
K}_{\alpha }^{\prime }+Q(\Phi ,\ul{\phi },K^{\prime }),
\label{cannonaback}
\end{equation}
where $Q$ is a $\ul{K}^{\prime }$- and $\ul{C}$-independent
local functional such that 
\begin{equation}
\llbracket \bar{S},Q(\Phi ,\ul{\phi },K)\rrbracket =0,
\label{assumback}
\end{equation}
satisfies our requirements.

Since $Q$ is $\ul{K}^{\prime }$ and $\ul{C}$ independent, the
background fields and the sources $\ul{K}_{C}$ do not transform: $%
\ul{\Phi }^{\prime }=\ul{\Phi }$, $\ul{K}_{C}^{\prime }=%
\ul{K}_{C}$. Moreover, the action $\hat{S}^{\prime }$ is clearly $%
\ul{K}^{\prime }$ and $\ul{C}^{\prime }$ independent, as
desired, so we just need to prove (\ref{thesisback}). For convenience,
multiply $Q$ by a constant parameter $\zeta $ and consider the canonical
transformations generated by 
\begin{equation}
F_{\zeta }(\Phi ,\ul{\Phi },K^{\prime },\ul{K}^{\prime })=\int
\ \Phi ^{\alpha }K_{\alpha }^{\prime }+\int \ \ul{\Phi }^{\alpha }%
\ul{K}_{\alpha }^{\prime }+\zeta Q(\Phi ,\ul{\phi },K^{\prime
}).  \label{fg}
\end{equation}
Given a functional $X(\Phi ,\ul{\Phi },K^{\prime },\ul{K}%
^{\prime })$ it is often useful to work with the tilde functional 
\begin{equation}
\tilde{X}(\Phi ,\ul{\Phi },K,\ul{K})=X(\Phi ,\ul{\Phi }%
,K^{\prime }(\Phi ,\ul{\Phi },K,\ul{K}),\ul{K}^{\prime
}(\Phi ,\ul{\Phi },K,\ul{K})).  \label{tildedfback}
\end{equation}
obtained by expressing the primed sources in terms of unprimed fields and
sources. Assumption (\ref{assumback}) tells us that $Q(\Phi ,\ul{\phi 
},K)$ is invariant under background transformations. Since $\Phi ^{\alpha }$
and $K_{\beta }$ transform as matter fields under such transformations, it
is clear that $\delta Q/\delta K_{\alpha }$ and $\delta Q/\delta \Phi
^{\beta }$ transform precisely like them, as well as $\Phi ^{\alpha \hspace{%
0.01in}\prime }$ and $K_{\beta }^{\prime }$. Moreover, we have $\llbracket 
\bar{S},\tilde{Q}\rrbracket =0$ for every $\zeta $. Applying theorem \ref
{theorem5} to $\chi =\bar{S}$ we obtain 
\begin{equation}
\frac{\partial ^{\prime }\bar{S}^{\prime }}{\partial \zeta }=\frac{\partial 
\bar{S}}{\delta \zeta }-\llbracket \bar{S},\tilde{Q}\rrbracket =\frac{%
\partial \bar{S}}{\delta \zeta },  \label{tocback}
\end{equation}
where $\partial ^{\prime }/\partial \zeta $ is taken at constant primed
variables and $\partial /\partial \zeta $ is taken at constant unprimed
variables. If we treat the unprimed variables as $\zeta $ independent, and
the primed variables as functions of them and $\zeta $, the right-hand side
of (\ref{tocback}) vanishes. Varying $\zeta $ from 0 to 1 we get 
\[
\bar{S}^{\prime }(\Phi ,\ul{\Phi },K,\ul{K})=\bar{S}^{\prime
}(\Phi ^{\prime },\ul{\Phi }^{\prime },K^{\prime },\ul{K}%
^{\prime })=\bar{S}(\Phi ,\ul{\Phi },K,\ul{K}), 
\]
where now the relations among primed and unprimed variables are those
specified by (\ref{cannonaback}).

We call the canonical transformations just defined \textit{%
background-preserving canonical transformations}. We stress once again that
they do not just preserve the background field ($\ul{\Phi }^{\prime }=%
\ul{\Phi }$), but also the background transformations ($\bar{S}%
^{\prime }=\bar{S}$) and the $\ul{K}$ and $\ul{C}$
independence of $\hat{S}$. The gauge-fixing canonical transformation (\ref
{backgfgen}) is background preserving.

Canonical transformations may convert the sources $K$ into functions of both
fields and sources. However, the sources are external, while the fields are
integrated over. Thus, canonical transformations must be applied at the
level of the action $S$, not at the levels of generating functionals. In the
functional integral they must be meant as mere replacements of
integrands. Nevertheless, we recall that there exists a way \cite{fieldcov,masterf,mastercan}
to upgrade the formalism of quantum field theory and overcome these
problems. The upgraded formalism allows us to implement canonical
transformations as true changes of field variables in the functional
integral, and closely track their effects inside generating functionals, as
well as throughout the renormalization algorithm.

\section{Renormalization}

\setcounter{equation}{0}

In this section we give the basic algorithm to subtract divergences to all
orders. As usual, we proceed by induction in the number of loops and use the
dimensional-regularization technique and the minimal subtraction scheme. We
assume that gauge anomalies are manifestly absent, i.e. that the
background field master equations (\ref{treide}) hold exactly at the
regularized level. We first work on the classical action $S=\hat{S}+\bar{S}$
of (\ref{sbaccagf}) and define a background-preserving subtraction
algorithm. Then we generalize the results to non-background-preserving
actions.

Call $S_{n}$ and $\Gamma _{n}$ the action and the $\Gamma $ functional
renormalized up to $n$ loops included, with $S_{0}=S$, and write the loop expansion as 
\[
\Gamma _{n}=\sum_{k=0}^{\infty }\hbar ^{n}\Gamma _{n}^{(k)}. 
\]
The inductive assumptions are that $S_{n}$ has the form (\ref{assu}), with $%
\bar{S}$ given by (\ref{sbar}), and
\begin{eqnarray}
S_{n} &=&S+\text{poles},\qquad \Gamma _{n}^{(k)}<\infty ~~\forall k\leqslant
n,  \label{assu1} \\
\llbracket S_{n},S_{n}\rrbracket &=&\mathcal{O}(\hbar ^{n+1}),\qquad %
\llbracket \bar{S},S_{n}\rrbracket =0,  \label{assu2}
\end{eqnarray}
where ``poles'' refers to the divergences of the dimensional regularization.
Clearly, the assumptions (\ref{assu1}) and (\ref{assu2}) are satisfied for $%
n=0$.

Using formulas (\ref{give}) and recalling that $\llbracket 
S_{n},S_{n}\rrbracket $ is a local insertion of order $\mathcal{O}(\hbar
^{n+1})$, we have 
\begin{equation}
\llbracket \Gamma _{n},\Gamma _{n}\rrbracket =\langle \llbracket S_{n},S_{n}%
\rrbracket \rangle =\llbracket S_{n},S_{n}\rrbracket +\mathcal{O}(\hbar
^{n+2}),\qquad \llbracket \bar{S},\Gamma _{n}\rrbracket =\langle \llbracket 
\bar{S},S_{n}\rrbracket \rangle =0.  \label{gnback2}
\end{equation}
By $\llbracket S,S\rrbracket =0$ and the first of (\ref{assu1}), $%
\llbracket 
S_{n},S_{n}\rrbracket $ is made of pure poles.

Now, take the order $\hbar ^{n+1}$ of equations (\ref{gnback2}) and then
their divergent parts. The second of (\ref{assu1}) tells us that all
subdivergences are subtracted away, so the order-$\hbar ^{n+1}$ divergent
part $\Gamma _{n\text{ div}}^{(n+1)}$ of $\Gamma _{n}$ is a local
functional. We obtain 
\begin{equation}
\llbracket S,\Gamma _{n\text{ div}}^{(n+1)}\rrbracket =\frac{1}{2}\llbracket %
S_{n},S_{n}\rrbracket +\mathcal{O}(\hbar ^{n+2}),\qquad \llbracket \bar{S}%
,\Gamma _{n\text{ div}}^{(n+1)}\rrbracket =0.  \label{gn2back}
\end{equation}

Define 
\begin{equation}
S_{n+1}=S_{n}-\Gamma _{n\text{ div}}^{(n+1)}.  \label{snp1back}
\end{equation}
Since $S_{n}$ has the form (\ref{assu}), $\Gamma _{n}$ has the
form (\ref{becco}), therefore both $\hat{\Gamma}_{n}$ and $\Gamma _{n\text{ div}%
}^{(n+1)}$ are $\ul{K}$ and $\ul{C}$ independent, which
ensures that $S_{n+1}$ has the form (\ref{assu}) (with $\bar{S}$ given by (%
\ref{sbar})). Moreover, the first inductive assumption of (\ref{assu1}) is
promoted to $S_{n+1}$. The diagrams constructed with the vertices of $%
S_{n+1} $ are the diagrams of $S_{n}$, plus new diagrams containing vertices
of $-\Gamma _{n\text{ div}}^{(n+1)}$; therefore 
\[
\Gamma _{n+1}^{(k)}=\Gamma _{n}^{(k)}<\infty ~~\forall k\leqslant n,\qquad
\Gamma _{n+1}^{(n+1)}=\Gamma _{n}^{(n+1)}-\Gamma _{n\text{ div}%
}^{(n+1)}<\infty , 
\]
which promotes the second inductive assumption of (\ref{assu1}) to $n+1$
loops. Finally, formulas (\ref{gn2back}) and (\ref{snp1back}) give 
\[
\llbracket S_{n+1},S_{n+1}\rrbracket =\llbracket S_{n},S_{n}\rrbracket -2%
\llbracket S,\Gamma _{n\text{ div}}^{(n+1)}\rrbracket +\mathcal{O}(\hbar
^{n+2})=\mathcal{O}(\hbar ^{n+2}),\qquad \llbracket \bar{S},S_{n+1}%
\rrbracket =0, 
\]
so (\ref{assu2}) are also promoted to $n+1$ loops.

We conclude that the renormalized action $S_{R}=S_{\infty }$ and the
renormalized generating functional $\Gamma _{R}=\Gamma _{\infty }$ satisfy
the background field master equations 
\begin{equation}
\llbracket S_{R},S_{R}\rrbracket =\llbracket \bar{S},S_{R}\rrbracket %
=0,\qquad \llbracket \Gamma _{R},\Gamma _{R}\rrbracket =\llbracket \bar{S}%
,\Gamma _{R}\rrbracket =0.  \label{finback}
\end{equation}
For later convenience we write down the form of $S_{R}$, which is 
\begin{equation}
S_{R}(\Phi ,\ul{\Phi },K,\ul{K})=\hat{S}_{R}(\Phi ,\ul{%
\phi },K)+\bar{S}(\Phi ,\ul{\Phi },K,\ul{K})=\hat{S}_{R}(\Phi ,%
\ul{\phi },K)-\int \mathcal{R}^{\alpha }(\Phi ,\ul{C}%
)K_{\alpha }-\int R^{\alpha }(\ul{\Phi })\ul{K}_{\alpha }.
\label{sr1}
\end{equation}

In the usual (non-background field) approach the results just derived hold
if we just ignore background fields and sources, as well as background
transformations, and use the standard parentheses $(X,Y)$ instead of $%
\llbracket X,Y\rrbracket $. Then the subtraction algorithm starts with a
classical action $S(\Phi ,K)$ that satisfies the usual master equation $%
(S,S)=0$ exactly at the regularized level and ends with a renormalized
action $S_{R}(\Phi ,K)=S_{\infty }(\Phi ,K)$ and a renormalized generating
functional $\Gamma _{R}(\Phi ,K)=\Gamma _{\infty }(\Phi ,K)$ that satisfy
the usual master equations $(S_{R},S_{R})=(\Gamma _{R},\Gamma _{R})=0$.

In the presence of background fields $\ul{\Phi }$ and background
sources $\ul{K}$, ignoring invariance under background
transformations (encoded in the parentheses $\llbracket \bar{S},S\rrbracket $%
, $\llbracket \bar{S},S_{n}\rrbracket $, $\llbracket \bar{S},S_{R}%
\rrbracket 
$ and similar ones for the $\Gamma $ functionals), we can generalize the
results found above to any classical action $S(\Phi ,\ul{\Phi },K,%
\ul{K})$ that satisfies $\llbracket S,S\rrbracket =0$ at the
regularized level and is such that $\delta _{l}S/\delta \ul{K}%
_{\alpha }$ is $\Phi $ independent. Indeed, these assumptions allow us to
apply theorem \ref{thb}, instead of formulas (\ref{give}), which is enough
to go through the subtraction algorithm ignoring the parentheses $%
\llbracket \bar{S},X\rrbracket $. We have $\delta _{l}S/\delta \ul{K}%
_{\alpha }=\delta _{l}\Gamma /\delta \ul{K}_{\alpha }=\delta
_{l}S_{n}/\delta \ul{K}_{\alpha }$ for every $n$. Thus, we conclude
that a classical action $S(\Phi ,\ul{\Phi },K,\ul{K})$ that
satisfies $\llbracket S,S\rrbracket =0$ at the regularized level and is such
that $\delta _{l}S/\delta \ul{K}_{\alpha }$ is $\Phi $ independent
gives a renormalized action $S_{R}(\Phi ,\ul{\Phi },K,\ul{K})$
and a $\Gamma $ functional $\Gamma _{R}(\Phi ,\ul{\Phi },K,\ul{%
K})$ that satisfy $\llbracket S_{R},S_{R}\rrbracket =\llbracket \Gamma
_{R},\Gamma _{R}\rrbracket =0$ and $\delta _{l}S_{R}/\delta \ul{K}%
_{\alpha }=\delta _{l}\Gamma _{R}/\delta \ul{K}_{\alpha }=\delta
_{l}S/\delta \ul{K}_{\alpha }$.

\medskip

The renormalization algorithm of this section is a generalization to the
background field method of the procedure first given in ref. \cite{lavrov}.
Since it subtracts divergences just as they come, as emphasized by formula (%
\ref{snp1back}), we use to call it ``raw'' subtraction \cite{regnocoho}, to
distinguish it from algorithms where divergences are subtracted away at each
step by means of parameter redefinitions and canonical transformations. The
raw subtraction does not ensure RG invariance \cite{regnocoho}, because it
subtracts divergent terms even when there is no (running) parameter
associated with them. For the same reason, it tells us very little about
parametric completeness.

In power-counting renormalizable theories the raw subtraction is
satisfactory, since we can start from a classical action $S_{c}$ that
already contains all gauge-invariant terms that are generated back by
renormalization. Nevertheless, in nonrenormalizable theories, such as
quantum gravity, effective field theories and nonrenormalizable extensions
of the Standard Model, in principle renormalization can modify the symmetry
transformations in physically observable ways (see ref. \cite{regnocoho} for
a discussion about this possibility). In section 5 we prove that this
actually does not happen under the assumptions we have made in this paper;
namely when gauge anomalies are manifestly absent, the gauge algebra is
irreducible and closes off shell, and $R^{\alpha }(\Phi )$ are quadratic
functions of the fields $\Phi $. Precisely, renormalization affects the
symmetry only by means of canonical transformations and parameter
redefinitions. Then, to achieve parametric completeness it is sufficient to
include all gauge-invariant terms in the classical action $S_{c}(\phi )$, as
classified by the starting gauge symmetry. The background field method is
crucial to prove this result without advocating involved cohomological
classifications.

\section{Gauge dependence}

\setcounter{equation}{0}

In this section we study the dependence on the gauge fixing and the
renormalization of canonical transformations. We first derive the
differential equations that govern gauge dependence; then we integrate them
and finally use the outcome to describe the renormalized canonical
transformation that switches between the background field approach and the
conventional approach. These results will be useful in the next section to
prove parametric completeness.

The parameters of a canonical transformation are associated with changes of
field variables and changes of gauge fixing. For brevity we call all of them
``gauge-fixing parameters'' and denote them with $\xi $. Let (\ref
{cannonaback}) be a tree-level canonical transformation satisfying (\ref
{assumback}). We write $Q(\Phi ,\ul{\phi },K^{\prime },\xi )$ to
emphasize the $\xi $ dependence of $Q$. We prove that for every gauge-fixing
parameter $\xi $ there exists a local $\ul{K}$- and $\ul{C}$-independent functional $Q_{R,\xi }$ such that 
\begin{equation}
Q_{R,\xi }=\widetilde{Q_{\xi }}+\mathcal{O}(\hbar )\text{-poles},\qquad
\langle Q_{R,\xi }\rangle <\infty ,  \label{babaoback}
\end{equation}
and
\begin{equation}
\frac{\partial S_{R}}{\partial \xi }=\llbracket S_{R},Q_{R,\xi }\rrbracket %
,\qquad \llbracket \bar{S},Q_{R,\xi }\rrbracket =0,\qquad \frac{\partial
\Gamma _{R}}{\partial \xi }=\llbracket \Gamma _{R},\langle Q_{R,\xi }\rangle %
\rrbracket ,  \label{backgind}
\end{equation}
where $Q_{\xi }=\partial Q/\partial \xi $, $\widetilde{Q_{\xi }}$ is defined
as shown in (\ref{tildedfback}) and the average is calculated with the
action $S_{R}$.
We call the first and last equations of the list (\ref{backgind}) \textit{differential
equations of} \textit{gauge dependence}. They ensure that 
renormalized functionals depend on gauge-fixing parameters in a cohomologically
exact way. Later we integrate equations (\ref{backgind}) and move every gauge
dependence inside a (renormalized) canonical transformation. A consequence
is that physical quantities are gauge independent.

We derive (\ref{backgind}) proceeding inductively in the number of loops, as
usual. The inductive assumption is that there exists a $\ul{K}$- and $%
\ul{C}$-independent local functional $Q_{n,\xi }=\widetilde{Q_{\xi }}+%
\mathcal{O}(\hbar )$-poles such that $\langle Q_{n,\xi }\rangle $ is
convergent up to the $n$th loop included (the average being calculated with
the action $S_{n}$) and 
\begin{equation}
\frac{\partial S_{n}}{\partial \xi }=\llbracket S_{n},Q_{n,\xi }\rrbracket +%
\mathcal{O}(\hbar ^{n+1}),\qquad \llbracket \bar{S},Q_{n,\xi }\rrbracket =0.
\label{inda1back}
\end{equation}
Applying the identity (\ref{thesis}), which here holds with the parentheses $%
\llbracket X,Y\rrbracket $, we easily see that $Q_{0,\xi }=\widetilde{Q_{\xi
}}$ satisfies (\ref{inda1back}) for $n=0$. Indeed, taking $\chi =S$ and
noting that $\left. \partial S^{\prime }/\partial \xi \right| _{\Phi
^{\prime },K^{\prime }}=0$, since the parameter $\xi $ is absent before the
transformation (a situation that we describe using primed variables), we get
the first relation of (\ref{inda1back}), without $\mathcal{O}(\hbar )$ corrections.
Applying (\ref{thesis}) to $\chi =\bar{S}$ and recalling that $\bar{S}$ is
invariant, we get the second relation of (\ref{inda1back}).

Let $Q_{n,\xi \hspace{0.01in}\text{div}}^{(n+1)}$ denote the $\mathcal{O}%
(\hbar ^{n+1})$ divergent part of $\langle Q_{n,\xi }\rangle $. The
inductive assumption ensures that all subdivergences are subtracted away, so 
$Q_{n,\xi \hspace{0.01in}\text{div}}^{(n+1)}$ is local. Define 
\begin{equation}
Q_{n+1,\xi }=Q_{n,\xi }-Q_{n,\xi \hspace{0.01in}\text{div}}^{(n+1)}.
\label{refdback}
\end{equation}
Clearly, $Q_{n+1,\xi }$ is $\ul{K}$ and $\ul{C}$ independent
and equal to $\widetilde{Q_{\xi }}+\mathcal{O}(\hbar )$-poles. Moreover, by
construction $\langle Q_{n+1,\xi }\rangle $ is convergent up to the $(n+1)$%
-th loop included, where the average is calculated with the action $%
S_{n+1}$.

Now, corollary \ref{corolla} tells us that $\llbracket 
\bar{S},Q_{n,\xi }\rrbracket =0$ and $\llbracket 
\bar{S},S_{n}\rrbracket =0$ imply $\llbracket 
\bar{S},\langle Q_{n,\xi }\rangle \rrbracket =0$. Taking the $\mathcal{O}%
(\hbar ^{n+1})$ divergent part of this formula we obtain $\llbracket 
\bar{S},Q_{n,\xi \hspace{0.01in}\text{div}}^{(n+1)}\rrbracket =0$; therefore
the second formula of (\ref{inda1back}) is promoted to $n+1$ loops.

Applying corollary \ref{cora} to $\Gamma _{n}$ and $S_{n}$, with $X=Q_{n,\xi
}$, we have the identity 
\begin{equation}
\frac{\partial \Gamma _{n}}{\partial \xi }=\llbracket \Gamma _{n},\langle
Q_{n,\xi }\rangle \rrbracket +\left\langle \frac{\partial S_{n}}{\partial
\xi }-\llbracket S_{n},Q_{n,\xi }\rrbracket \right\rangle +\frac{1}{2}%
\left\langle \llbracket S_{n},S_{n}\rrbracket \hspace{0.01in}Q_{n,\xi
}\right\rangle _{\Gamma },  \label{provef}
\end{equation}
where $\left\langle AB\right\rangle _{\Gamma }$ denotes the one-particle
irreducible diagrams with one $A$ insertion and one $B$ insertion. Now,
observe that if $A=\mathcal{O}(\hbar ^{n_{A}})$ and $B=\mathcal{O}(\hbar
^{n_{B}})$ then $\left\langle AB\right\rangle _{\Gamma }=\mathcal{O}(\hbar
^{n_{A}+n_{B}+1})$, since the $A,B$ insertions can be connected only by
loops. Let us take the $\mathcal{O}(\hbar ^{n+1})$ divergent part of (\ref
{provef}). By the inductive assumption (\ref{assu2}), the last term of (\ref
{provef}) can be neglected. By the inductive assumption (\ref{inda1back}) we
can drop the average in the second-to-last term. We thus get 
\[
\frac{\partial \Gamma _{n\ \text{div}}^{(n+1)}}{\partial \xi }=\llbracket 
\Gamma _{n\ \text{div}}^{(n+1)},Q_{0,\xi }\rrbracket +\llbracket 
S,Q_{n,\xi \hspace{0.01in}\text{div}}^{(n+1)}\rrbracket +\frac{\partial S_{n}%
}{\partial \xi }-\llbracket 
S_{n},Q_{n,\xi }\rrbracket +\mathcal{O}(\hbar ^{n+2}). 
\]
Using this fact, (\ref{snp1back}) and (\ref{refdback}) we obtain 
\begin{equation}
\frac{\partial S_{n+1}}{\partial \xi }=\llbracket S_{n+1},Q_{n+1,\xi }%
\rrbracket +\mathcal{O}(\hbar ^{n+2}),  \label{sunpi}
\end{equation}
which promotes the first inductive hypothesis of (\ref{inda1back}) to order $%
\hbar ^{n+1}$. When $n$ is taken to infinity, the first two formulas of (\ref
{backgind}) follow, with $Q_{R,\xi }=Q_{\infty ,\xi }$. The third identity
of (\ref{backgind}) follows from the first one, using (\ref{provef}) with $%
n=\infty $ and $\llbracket 
\hat{S}_{R},\hat{S}_{R}\rrbracket =0$. This concludes the derivation of (\ref
{backgind}).

\subsection{Integrating the differential equations of gauge dependence}
\label{integrating}

Now we integrate the first two equations of (\ref{backgind}) and find the renormalized
canonical transformation that corresponds to a tree-level transformation (%
\ref{cannonaback}) satisfying (\ref{assumback}). Specifically, we prove that

\begin{theorem}
There exists a background-preserving canonical transformation 
\begin{equation}
F_{R}(\Phi ,\ul{\Phi },K^{\prime },\ul{K}^{\prime },\xi )=\int
\ \Phi ^{A}K_{A}^{\prime }+\int \ \ul{\Phi }^{A}\ul{K}%
_{A}^{\prime }+Q_{R}(\Phi ,\ul{\phi },K^{\prime },\xi ),
\label{finalcanback}
\end{equation}
where $Q_{R}(\Phi ,\ul{\phi },K^{\prime },\xi )=Q(\Phi ,\ul{%
\phi },K^{\prime },\xi )+\mathcal{O}(\hbar )$ is a $\ul{K}$- and $%
\ul{C}$-independent local functional, such that the transformed
action $S_{f}(\Phi ^{\prime },\ul{\Phi }^{\prime },K^{\prime },%
\ul{K}^{\prime })=S_{R}(\Phi ,\ul{\Phi },K,\ul{K},\xi )$
is $\xi $ independent and invariant under background transformations: 
\begin{equation}
\frac{\partial S_{f}}{\partial \xi }=0,\qquad \llbracket \bar{S},S_{%
f}\rrbracket =0.  \label{gindep2back}
\end{equation}
\end{theorem}
\textit{Proof}. To prove this statement we introduce a new parameter $\zeta $
multiplying the whole functional $Q$ of (\ref{cannonaback}), as in (\ref{fg}%
). We know that $\llbracket \bar{S},Q\rrbracket =0$ implies $\llbracket \bar{%
S},\tilde{Q}\rrbracket =0$. If we prove that the $\zeta $ dependence can be
reabsorbed into a background-preserving canonical transformation we also
prove the same result for every gauge-fixing parameter $\xi $ and also for all of them together. The
differential equations of gauge dependence found above obviously apply with $%
\xi \rightarrow \zeta $.

Specifically, we show that the $\zeta $ dependence can be reabsorbed in a
sequence of background-preserving canonical transformations $S_{R\hspace{%
0.01in}n}\rightarrow S_{R\hspace{0.01in}n+1}$ (with $S_{R\hspace{0.01in}%
0}=S_{R}$), generated by 
\begin{equation}
F_{n}(\Phi ,\ul{\Phi },K^{\prime },\ul{K}^{\prime })=\int \
\Phi ^{A}K_{A}^{\prime }+\int \ \ul{\Phi }^{A}\ul{K}%
_{A}^{\prime }+H_{n}(\Phi ,\ul{\phi },K^{\prime },\zeta ),  \label{fn}
\end{equation}
where $H_{n}=\mathcal{O}(\hbar ^{n})$, and such that 
\begin{equation}
\frac{\partial S_{R\hspace{0.01in}n}}{\partial \zeta }=\llbracket S_{R%
\hspace{0.01in}n},T_{n}\rrbracket ,\qquad T_{n}=\mathcal{O}(\hbar ^{n}).
\label{tn}
\end{equation}
The functionals $T_{n}$ and $H_{n}$ are determined by the recursive
relations 
\begin{eqnarray}
T_{n+1}(\Phi ^{\prime },\ul{\phi },K^{\prime },\zeta ) &=&T_{n}(\Phi ,%
\ul{\phi },K,\zeta )-\widetilde{\frac{\partial H_{n}}{\partial \zeta }%
},  \label{d1} \\
H_{n}(\Phi ,\ul{\phi },K^{\prime },\zeta ) &=&\int_{0}^{\zeta }d\zeta
^{\prime }\hspace{0.01in}T_{n,n}(\Phi ,\ul{\phi },K^{\prime },\zeta
^{\prime }),  \label{d2}
\end{eqnarray}
with the initial conditions 
\[
T_{0}=Q_{R,\zeta },\qquad H_{0}=\zeta Q. 
\]
In formula (\ref{d1}) the tilde operation (\ref{tildedfback}) on $\partial
H_{n}/\partial \zeta $ and the canonical transformation $\Phi ,K\rightarrow
\Phi ^{\prime },K^{\prime }$ are the ones defined by $F_{n}$. In formula (%
\ref{d2}) $T_{n,n}(\Phi ,\ul{\phi },K^{\prime })$ denotes the
contributions of order $\hbar ^{n}$ to $T_{n}(\Phi ,\ul{\phi },K(\Phi
,\ul{\phi },K^{\prime }))$, the function $K(\Phi ,\ul{\phi }%
,K^{\prime })$ also being determined by $F_{n}$. Note that for $n>0$ we have 
$T_{n}(\Phi ,\ul{\phi },K(\Phi ,\ul{\phi },K^{\prime
}))=T_{n}(\Phi ,\ul{\phi },K^{\prime })+\mathcal{O}(\hbar ^{n+1})$,
therefore formula (\ref{d2}), which determines $H_{n}$ (and so $F_{n}$),
does not really need $F_{n}$ on the right-hand side. Finally, (\ref{d2}) is
self-consistent for $n=0$.

Formula (\ref{thesis}) of the appendix describes how the dependence on
parameters is modified by a canonical transformation. Applying it to (\ref
{tn}), we get 
\[
\frac{\partial S_{R\hspace{0.01in}n+1}}{\partial \zeta }=\frac{\partial S_{R%
\hspace{0.01in}n}}{\partial \zeta }-\llbracket S_{R\hspace{0.01in}n},%
\widetilde{\frac{\partial H_{n}}{\partial \zeta }}\rrbracket =\llbracket S_{R%
\hspace{0.01in}n},T_{n}-\widetilde{\frac{\partial H_{n}}{\partial \zeta }}%
\rrbracket , 
\]
whence (\ref{d1}) follows. For $n=0$ the first formula of (\ref{babaoback})
gives $T_{0}=\widetilde{Q}+\mathcal{O}(\hbar )$, therefore $T_{1}=\mathcal{O}%
(\hbar )$. Then (\ref{d2}) gives $H_{1}=\mathcal{O}(\hbar )$. For $n>0$ the
order $\hbar ^{n}$ of $T_{n+1}$ vanishes by formula (\ref{d2}); therefore $%
T_{n+1}=\mathcal{O}(\hbar ^{n+1})$ and $H_{n+1}=\mathcal{O}(\hbar ^{n+1})$,
as desired.

Consequently, $S_{f}\equiv S_{R\hspace{0.01in}\infty }$ is $\zeta $
 independent, since (\ref{tn}) implies $\partial S_{R\hspace{0.01in}\infty
}/\partial \zeta =0$. Observe that $\ul{K}$ and $\ul{C}$
independence is preserved at each step. Finally, all operations defined by (%
\ref{d1}) and (\ref{d2}) are background preserving. We conclude that the
canonical transformation $F_{R}$ obtained composing the $F_{n}$s solves the
problem.

Using (\ref{gindep2back}) and (\ref{give}) we conclude that in the new
variables 
\begin{equation}
\frac{\partial \Gamma _{f}}{\partial \xi }=\left\langle \frac{%
\partial S_{f}}{\partial \xi }\right\rangle =0\qquad \llbracket \bar{S%
},\Gamma _{f}\rrbracket =0,  \label{finalmenteback}
\end{equation}
for all gauge-fixing parameters $\xi $.

\subsection{Non-background-preserving canonical transformations}

In the usual approach the results derived so far apply with straightforward
modifications. It is sufficient to ignore the background fields and sources,
as well as the background transformations, and use the standard parentheses $%
(X,Y)$ instead of $\llbracket X,Y\rrbracket $. Thus, given a tree-level
canonical transformation generated by 
\begin{equation}
F(\Phi ,K^{\prime })=\int \ \Phi ^{\alpha }K_{\alpha }^{\prime }+Q(\Phi
,K^{\prime },\xi ),  \label{f1}
\end{equation}
there exists a local functional $Q_{R,\xi }$ satisfying (\ref{babaoback})
such that 
\begin{equation}
\frac{\partial S_{R}}{\partial \xi }=(S_{R},Q_{R,\xi }),\qquad \frac{%
\partial \Gamma _{R}}{\partial \xi }=(\Gamma _{R},\langle Q_{R,\xi }\rangle
),  \label{br}
\end{equation}
and there exists a renormalized canonical transformation 
\begin{equation}
F_{R}(\Phi ,K^{\prime })=\int \ \Phi ^{A}K_{A}^{\prime }+Q_{R}(\Phi
,K^{\prime },\xi ),  \label{f2}
\end{equation}
where $Q_{R}(\Phi ,K^{\prime },\xi )=Q(\Phi ,K^{\prime },\xi )+\mathcal{O}%
(\hbar )$ is a local functional, such that the transformed action $S_{f
}(\Phi ^{\prime },K^{\prime })=S_{R}(\Phi ,K,\xi )$ is $\xi $ independent.
Said differently, the entire $\xi $ dependence of $S_{R}$ is reabsorbed into
the transformation: 
\[
S_{R}(\Phi ,K,\xi )=S_{f}(\Phi ^{\prime }(\Phi ,K,\xi ),K^{\prime
}(\Phi ,K,\xi )). 
\]

\medskip

In the presence of background fields $\ul{\Phi }$ and background
sources $\ul{K}$, dropping assumption (\ref{assumback}) and ignoring
invariance under background transformations, encoded in the parentheses $%
\llbracket \bar{S},X\rrbracket $, the results found above can be easily
generalized to any classical action $S(\Phi ,\ul{\Phi },K,\ul{K%
})$ that solves $\llbracket S,S\rrbracket =0$ and is such that $\delta
_{l}S/\delta \ul{K}_{\alpha }$ is $\Phi $ independent, and to any $%
\ul{K}$-independent canonical transformation. Indeed, these
assumptions are enough to apply theorem \ref{thb} and corollary \ref{cora},
and go through the derivation ignoring the parentheses $%
\llbracket \bar{S},X\rrbracket $. The tree-level canonical transformation is
described by a generating functional of the form 
\begin{equation}
F(\Phi ,\ul{\Phi },K^{\prime },\ul{K}^{\prime })=\int \ \Phi
^{\alpha }K_{\alpha }^{\prime }+\int \ \ul{\Phi }^{\alpha }\ul{%
K}_{\alpha }^{\prime }+Q(\Phi ,\ul{\phi },\ul{C},K^{\prime
},\xi ).  \label{fa1}
\end{equation}
We still find the differential equations 
\begin{equation}
\frac{\partial S_{R}}{\partial \xi }=\llbracket S_{R},Q_{R,\xi }\rrbracket %
,\qquad \frac{\partial \Gamma _{R}}{\partial \xi }=\llbracket \Gamma
_{R},\langle Q_{R,\xi }\rangle \rrbracket ,  \label{eqw}
\end{equation}
where $Q_{R,\xi }$ satisfies (\ref{babaoback}). When we integrate the first of these
equations with the procedure defined above we build a renormalized canonical
transformation 
\begin{equation}
F_{R}(\Phi ,\ul{\Phi },K^{\prime },\ul{K}^{\prime },\xi )=\int
\ \Phi ^{\alpha}K_{\alpha}^{\prime }+\int \ \ul{\Phi }^{\alpha}\ul{K}%
_{\alpha}^{\prime }+Q_{R}(\Phi ,\ul{\phi },\ul{C},K^{\prime },\xi ),
\label{fra1}
\end{equation}
where $Q_{R}=Q+\mathcal{O}(\hbar )$ is a local functional, such that the transformed action $S_{%
f}(\Phi ^{\prime },\ul{\Phi }^{\prime },K^{\prime },\ul{%
K}^{\prime })=S_{R}(\Phi ,\ul{\Phi },K,\ul{K},\xi )$ is $\xi $%
 independent. The only difference is that now $Q_{R,\xi }$, $\langle
Q_{R,\xi }\rangle $, $T_{n}$, $H_{n}$ and $Q_{R}$ can
depend on $\ul{C}$, which does not disturb any of the arguments used
in the derivation.

\subsection{Canonical transformations in \texorpdfstring{$\Gamma$}{}}

We have integrated the first equation of (\ref{eqw}), and shown that the $\xi$ dependence can be reabsorbed in the canonical transformation (\ref{fra1}) on the renormalized action $S_{R}$, which gives the $\xi$-independent action with $S_{\rm f}$. We know that the generating functional $\Gamma_{\rm f}$ of one-particle irreducible Green functions determined by $S_{\rm f}$ is $\xi$ independent. We can also prove that $\Gamma_{\rm f}$ can be obtained applying a (non-local) canonical transformation directly on $\Gamma_{R}$.

To achieve this goal we integrate the second equation of (\ref{eqw}). The integration algorithm is the same as the one of subsection \ref{integrating}, with the difference that $Q_{R,\xi}$ is replaced by $\langle Q_{R,\xi}\rangle$. The canonical transformation on $\Gamma_{R}$ has a generating functional of the form
\begin{equation}
F_{\Gamma}(\Phi ,\ul{\Phi },K^{\prime },\ul{K}^{\prime },\xi )=\int
\ \Phi ^{\alpha}K_{\alpha}^{\prime }+\int \ \ul{\Phi }^{\alpha}\ul{K}%
_{\alpha}^{\prime }+Q_{\Gamma}(\Phi ,\ul{\phi },\ul{C},K^{\prime },\xi ),
\label{fra2}
\end{equation}
where $Q_{\Gamma}=Q+{\cal O}(\hbar)$ (non-local) radiative corrections. 

The result just obtained is actually more general, and proves that if $S$ is any action that solves the master equation (it can be the classical action, the renormalized action, or any other action) canonical transformations on $S$ correspond to canonical transformations on the $\Gamma$ functional determined by $S$. See \cite{quadri} for a different derivation of this result in Yang-Mills theory. Our line of reasoning can be recapitulated as follows: in the usual approach, ($i$) make a canonical transformation (\ref{f1}) on $S$; ($ii$) derive the equations of gauge dependence for the action, which are $\partial S/\partial\xi=( S,Q_{\xi}) $; ($iii$) derive the equations of gauge dependence for the $\Gamma$ functional determined by $S$, which are $\partial \Gamma/\partial\xi=( \Gamma,\langle Q_{\xi}\rangle)$, and integrate them.

The property just mentioned may sound obvious, and is often taken for granted, but actually needed to be proved. The reason is that the canonical transformations we are talking about are not true changes of field variables inside functional integrals, but mere replacements of integrands \cite{fieldcov}. Therefore, we cannot automatically infer how a transformation on the action $S$ affects the generating functionals $Z$, $W=\ln Z$ and $\Gamma$, and need to make some additional effort to get where we want. We recall that to skip this kind of supplementary analysis we need to use the formalism of the master functional, explained in refs. \cite{masterf,mastercan}. 

\subsection{Application}

An interesting application that illustrates the results of this section is
the comparison between the renormalized action (\ref{sr1}), which was
obtained with the background field method and the raw subtraction procedure
of section 3, and the renormalized action $S_{R}^{\prime }$ that can be
obtained with the same raw subtraction in the usual non-background field
approach.

The usual approach is retrieved by picking a gauge fermion $\Psi ^{\prime }$
that depends on $\Phi +\ul{\Phi }$, such as 
\begin{equation}
\Psi ^{\prime }(\Phi +\ul{\Phi })=\int \bar{C}^{I}G^{Ii}(0,\partial
)(\phi ^{i}+\ul{\phi }^{i}).  \label{gfno}
\end{equation}
Making the canonical transformation generated by 
\begin{equation}
F_{\text{gf}}^{\prime }(\Phi ,\ul{\Phi },K^{\prime },\ul{K}%
^{\prime })=\int \ \Phi ^{\alpha }K_{\alpha }^{\prime }+\int \ \ul{%
\Phi }^{\alpha }\ul{K}_{\alpha }^{\prime }+\Psi ^{\prime }(\Phi +%
\ul{\Phi })  \label{backgfgen2}
\end{equation}
on (\ref{sback}) we find the classical action 
\begin{equation}
S^{\prime }(\Phi ,\ul{\Phi },K,\ul{K})=\hat{S}^{\prime }(\Phi +%
\ul{\Phi },K)+\bar{S}^{\prime }(\Phi ,\ul{\Phi },K,\ul{K%
}),  \label{sr2c}
\end{equation}
where 
\begin{equation}
\hat{S}^{\prime }(\Phi +\ul{\Phi },K)=S_{c}(\phi +\ul{\phi }%
)-\int R^{\alpha }(\Phi +\ul{\Phi })\bar{K}_{\alpha },\qquad \bar{S}%
^{\prime }(\Phi ,\ul{\Phi },K,\ul{K})=-\int R^{\alpha }(%
\ul{\Phi })(\ul{K}_{\alpha }-K_{\alpha }),  \label{sr2gf}
\end{equation}
and the barred sources $\bar{K}_{\alpha }$ coincide with $K_{\alpha }$ apart
from $\bar{K}_{\phi }^{i}$ and $\bar{K}_{\bar{C}}^{I}$, which are 
\begin{equation}
\bar{K}_{\phi }^{i}=K_{\phi }^{i}-\bar{C}^{I}G^{Ii}(0,-\overleftarrow{%
\partial }),\qquad \bar{K}_{\bar{C}}^{I}=K_{\bar{C}}^{I}-G^{Ii}(0,\partial
)(\phi ^{i}+\ul{\phi }^{i}).  \label{chif2}
\end{equation}
Clearly, $\hat{S}^{\prime }$ is the gauge-fixed classical action of the
usual approach, apart from the shift $\Phi \rightarrow \Phi +\ul{\Phi 
}$.

The radiative corrections are generated only by $\hat{S}^{\prime }$ and do
not affect $\bar{S}^{\prime }$. Indeed, $\hat{S}^{\prime }$ as well as the
radiative corrections are unaffected by setting $\ul{\Phi }=0$ and then
shifting $\Phi $ back to $\Phi +\ul{\Phi }$, while $\bar{S}^{\prime }$
disappears doing this. Thus, $\bar{S}^{\prime }$ is nonrenormalized, and
the renormalized action $S_{R}^{\prime }$ has the form 
\begin{equation}
S_{R}^{\prime }(\Phi ,\ul{\Phi },K,\ul{K})=\hat{S}_{R}^{\prime
}(\Phi +\ul{\Phi },K)+\bar{S}^{\prime }(\Phi ,\ul{\Phi },K,%
\ul{K}).  \label{sr2}
\end{equation}

Now we compare the classical action (\ref{sbaccagf}) of the background field
method with the classical action (\ref{sr2c}) of the usual approach. We
recapitulate how they are obtained with the help of the following schemes:
\[
\begin{tabular}{cccccccccc}
(\ref{sback}) & $\stackrel{(\ref{casbacca})}{\mathrel{\scalebox{2.5}[1]{$%
\longrightarrow$}}}$ & (\ref{sbacca}) & $\stackrel{(\ref{backgfgen})}{%
\mathrel{\scalebox{2.5}[1]{$\longrightarrow$}}}$ & $S=$ (\ref{sbaccagf}) & 
\qquad \qquad & (\ref{sback}) & $\stackrel{(\ref{backgfgen2})}{%
\mathrel{\scalebox{2.5}[1]{$ \longrightarrow$}}}$ & $S^{\prime }=$ (\ref
{sr2gf}). & 
\end{tabular}
\]
Above the arrows we have put references to the corresponding canonical
transformations, which are (\ref{casbacca}), (\ref{backgfgen}) and (\ref
{backgfgen2}) and commute with one another. We can interpolate between the
classical actions (\ref{sbaccagf}) and (\ref{sr2gf})\ by means of a $%
\ul{K}$-independent non-background-preserving canonical
transformation generated by 
\begin{equation}
F_{\xi }(\Phi ,\ul{\Phi },K^{\prime },\ul{K}^{\prime },\xi
)=\int \ \Phi ^{\alpha }K_{\alpha }^{\prime }+\int \ \ul{\Phi }%
^{\alpha }\ul{K}_{\alpha }^{\prime }+\xi \Delta \Psi +\xi \int 
\mathcal{R}_{\bar{C}}^{I}(\bar{C},\ul{C})K_{B}^{I\hspace{0.01in}%
\prime }.  \label{ds}
\end{equation}
where $\xi $ is a gauge-fixing parameter that varies from 0 to 1, and 
\[
\Delta \Psi =\int \bar{C}^{I}\left( G^{Ii}(\ul{\phi },\partial )\phi
^{i}-G^{Ii}(0,\partial )(\phi ^{i}+\ul{\phi }^{i})\right) . 
\]

Precisely, start from the non-background field theory (\ref{sr2c}), and\
take its variables to be primed ones. We know that $\hat{S}^{\prime }$
depends on the combination 
\begin{equation}
\tilde{K}_{\phi }^{i\hspace{0.01in}\prime }=K_{\phi }^{i\hspace{0.01in}%
\prime }-\bar{C}^{I\hspace{0.01in}}G^{Ii}(0,-\overleftarrow{\partial }),
\label{kip}
\end{equation}
and we have $\bar{C}^{I\hspace{0.01in}}=\bar{C}^{I\hspace{%
0.01in}\prime }$. Expressing the primed fields and sources in terms of the
unprimed ones and $\xi $, we find the interpolating classical action 
\begin{equation}
S_{\xi }(\Phi ,\ul{\Phi },K,\ul{K})=S_{c}(\phi +\ul{%
\phi })-\int R^{\alpha }(\Phi +\ul{\Phi })\tilde{K}_{\alpha }(\xi
)-\xi \int \mathcal{R}_{\bar{C}}^{I}(\bar{C},\ul{C})\tilde{K}_{\bar{C}%
}^{I}(\xi )-\int R^{\alpha }(\ul{\Phi })(\ul{\tilde{K}}%
_{\alpha }(\xi )-\tilde{K}_{\alpha }(\xi )),  \label{sx}
\end{equation}
where $\tilde{K}_{C}^{I}(\xi )=K_{C}^{I}$, 
\begin{equation}
\tilde{K}_{\phi }^{i}(\xi )=K_{\phi }^{i}-\xi \bar{C}^{I\hspace{0.01in}%
}G^{Ii}(\ul{\phi },-\overleftarrow{\partial })-(1-\xi )\bar{C}^{I%
\hspace{0.01in}}G^{Ii}(0,-\overleftarrow{\partial }),  \label{convex}
\end{equation}
while the other $\xi $-dependent tilde sources have expressions that we do
not need to report here. It suffices to say that they are $K_{\phi }^{i}$ 
independent, such that $\delta _{r}S_{\xi }/\delta \ul{K}_{\alpha
}=-R^{\alpha }(\ul{\Phi })$, and linear in the quantum fields $\Phi $%
, apart from $\ul{\tilde{K}}_{\phi }^{i}(\xi )$, which is quadratic.
Thus the action $S_{\xi }$ and the transformation $F_{\xi }$ satisfy the
assumptions that allow us to apply theorem \ref{thb} and corollary \ref{cora}%
. Actually, (\ref{ds}) is of type (\ref{fa1}); therefore we have the
differential equations (\ref{eqw}) and the renormalized canonical
transformation (\ref{fra1}).

We want to better characterize the renormalized version $F_{R}$ of $F_{\xi }$%
. We know that the derivative of the renormalized $\Gamma $ functional with
respect to $\xi $ is governed by the renormalized version of the average 
\[
\left\langle \widetilde{\frac{\partial F_{\xi }}{\partial \xi }}%
\right\rangle =\langle \Delta \Psi \rangle +\int \mathcal{R}_{\bar{C}}^{I}(%
\bar{C},\ul{C})K_{B}^{I}. 
\]
It is easy to see that $\langle \Delta \Psi \rangle $ is independent of $%
\ul{K}$, $B$, $K_{\bar{C}}$ and $K_{B}$, since no one-particle
irreducible diagrams with such external legs can be constructed. In
particular, $K_{B}^{I\hspace{0.01in}\prime }=K_{B}^{I}$. Moreover, using the
explicit form (\ref{sx}) of the action $S_{\xi }$ and arguments similar to
the ones that lead to formulas (\ref{chif}), we easily see that $\langle
\Delta \Psi \rangle $ is equal to $\Delta \Psi $ plus a functional that does
not depend on $K_{\phi }^{i}$ and $\bar{C}^{I}$ separately,
but only on the convex combination (\ref{convex}). Indeed, the $\bar{C}$%
-dependent terms of (\ref{sx}) that do not fit into the combination (\ref
{convex}) are $K_{\phi }^{i}$ independent and at most quadratic in the
quantum fields, so they cannot generate one-particle irreducible diagrams
that have either $K_{\phi }^{i}$ or $\bar{C}$ on the external legs. Clearly,
the renormalization of $\langle \Delta \Psi \rangle $ also satisfies the
properties just stated for $\langle \Delta \Psi \rangle $.

Following the steps of the previous section we can integrate the $\xi $
derivative and reconstruct the full canonical transformation. However,
formula (\ref{d2}) shows that the integration over $\xi $ must be done
by keeping fixed the unprimed fields $\Phi $ and the primed sources $K^{\prime
} $. When we do this for the zeroth canonical transformation $F_{0}$ of (\ref
{fn}), the combination $\tilde{K}_{\phi }^{i}(\xi )$ is turned into (\ref
{kip}), which is $\xi $ independent. Every other transformation $F_{n}$ of (%
\ref{fn}) preserves the combination (\ref{kip}), so the integrated canonical
transformation does not depend on $K_{\phi }^{i}$ and $\bar{C}^{I}$ separately, but only on the combination $\tilde{K}_{\phi }^{i%
\hspace{0.01in}\prime }$, and the generating functional of the renormalized
version $F_{R}$ of $F_{\xi }$ has the form 
\begin{equation}
F_{R}(\Phi ,\ul{\Phi },K^{\prime },\ul{K}^{\prime },\xi )=\int
\ \Phi ^{\alpha }K_{\alpha }^{\prime }+\int \ \ul{\Phi }^{\alpha }%
\ul{K}_{\alpha }^{\prime }+\xi \Delta \Psi +\xi \int \mathcal{R}_{%
\bar{C}}^{I}(\bar{C},\ul{C})K_{B}^{I\hspace{0.01in}\prime }+\Delta
F_{\xi }(\phi ,C,\ul{\phi },\ul{C},\tilde{K}_{\phi }^{\prime
},K_{C}^{\prime },\xi ).  \label{fr}
\end{equation}
Using this expression we can verify {\it a posteriori} that indeed $\tilde{K}%
_{\phi }^{i}(\xi )$ depends just on (\ref{kip}), not on $K_{\phi }^{i\hspace{0.01in}\prime }$
and $\bar{C}^{I}$ separately. Moreover, (\ref{fr}) implies 
\begin{equation}
\ul{\Phi }^{\alpha \hspace{0.01in}\prime }=\ul{\Phi }^{\alpha},\qquad B^{I\hspace{0.01in}\prime }=B^{I}+\xi 
\mathcal{R}_{\bar{C}}^{I}(\bar{C},\ul{C}),\qquad \bar{C}^{I\hspace{0.01in}\prime }=%
\bar{C}^{I},\qquad K_{B}^{I\hspace{0.01in}\prime }=K_{B}^{I}.  \label{besid}
\end{equation}

In the next section these results are used to achieve parametric
completeness.

\section{Renormalization and parametric completeness}

\setcounter{equation}{0}

The raw renormalization algorithm of section 3 subtracts away divergences
just as they come. It does not ensure, \textit{per se}, RG invariance, for
which it is necessary to prove parametric completeness, namely that all
divergences can be subtracted by redefining parameters and making
canonical transformations.
We must show that we can include in the
classical action all invariants that are generated back by renormalization,
and associate an independent parameter with each of them. 
The purpose of this section is to show that the
background field method allows us to prove parametric completeness in a
rather straightforward way, making cohomological classifications unnecessary.

We want to relate the renormalized actions (\ref{sr1}) and (\ref{sr2}). From
the arguments of the previous section we know that these two actions are
related by the canonical transformation generated by (\ref{fr}) at $\xi =1$.
We have 
\begin{equation}
\hat{S}_{R}^{\prime }(\Phi ^{\prime }+\ul{\Phi },K^{\prime })-\int
R^{\alpha }(\ul{\Phi })(\ul{K}_{\alpha }^{\prime }-K_{\alpha
}^{\prime })=\hat{S}_{R}(\Phi ,\ul{\phi },K)-\int \mathcal{R}^{\alpha
}(\Phi ,\ul{C})K_{\alpha }-\int R^{\alpha }(\ul{\Phi })%
\ul{K}_{\alpha }.  \label{ei}
\end{equation}
From (\ref{fr}) we find the transformation rules (\ref{besid}) at $\xi =1$ and
\begin{equation}
\phi ^{\prime }=\phi +\frac{\delta \Delta F}{\delta \tilde{K}_{\phi
}^{\prime }},\qquad K_{\bar{C}}^{I}=K_{\bar{C}}^{I\hspace{0.01in}\prime
}+G^{Ii}(\ul{\phi },\partial )\phi ^{i}-G^{Ii}(0,\partial )(\phi ^{i%
\hspace{0.01in}\prime }+\ul{\phi }^{i})+\frac{\delta }{\delta \bar{C}%
^{I}}\int \mathcal{R}_{\bar{C}}^{J}(\bar{C},\ul{C})K_{B}^{J},
\label{beside}
\end{equation}
where $\Delta F$ is $\Delta F_{\xi}$ at $\xi=1$. Here and below we sometimes understand indices when there is no loss of clarity.
We want to express equation (\ref{ei}) in
terms of unprimed fields and primed sources, and then set $\phi =C=%
\ul{K}^{\prime }=0$. We denote this operation with a subscript 0.
Keeping $\ul{\phi },\ul{C},\bar{C},B$ and $K^{\prime }$ as
independent variables, we get 
\begin{eqnarray}
\hat{S}_{R}^{\prime }(\Phi _{0}^{\prime }+\ul{\Phi },K^{\prime }) &=&%
\hat{S}_{R}(0,\ul{\phi },0)-\int B^{I\prime }(K_{\bar{C}}^{I\hspace{%
0.01in}\prime }-G^{Ii}(0,\partial )(\phi _{0}^{i\hspace{0.01in}\prime }+%
\ul{\phi }^{i}))  \nonumber \\
&&-\int \mathcal{R}_{\bar{C}}^{I}(\bar{C},\ul{C})\frac{\delta }{%
\delta \bar{C}^{I}}\int \mathcal{R}_{\bar{C}}^{J}(\bar{C},\ul{C}%
)K_{B}^{J\hspace{0.01in}\prime }-\int R^{\alpha }(\ul{\Phi })(%
\ul{K}_{\alpha 0}+K_{\alpha }^{\prime }).  \label{uio}
\end{eqnarray}
To derive this formula we have used 
\begin{equation}
\hat{S}_{R}(\{0,0,\bar{C},B\},\ul{\phi },K)=\hat{S}_{R}(0,\ul{%
\phi },0)-\int B^{I}K_{\bar{C}}^{I},  \label{mina}
\end{equation}
together with (\ref{iddo}), (\ref{besid}) and (\ref{beside}). The reason why
(\ref{mina}) holds is that at $C=0$ there are no objects with positive ghost
numbers inside the left-hand side of this equation; therefore we can drop
every object that has a negative ghost number, which means $\bar{C}$ and all
sources $K$ but $K_{\bar{C}}^{I}$. Since (\ref{chio}) are the only $B$ and $%
K_{\bar{C}}^{I}$-dependent terms, and they are not renormalized, at $\phi =0$
we find (\ref{mina}).

Now, consider the canonical transformation $\{\ul{\phi },\ul{C}%
,\bar{C},B\},\breve{K}\rightarrow \Phi ^{\prime \prime },K^{\prime }$
defined by the generating functional 
\begin{equation}
F(\{\ul{\phi },\ul{C},\bar{C},B\},K^{\prime })=\int \ul{%
\phi }K_{\phi }^{\prime }+\int \ \ul{C}K_{C}^{\prime }+F_{R}(\{0,0,%
\bar{C},B\},\ul{\Phi },K^{\prime },0,1).  \label{for}
\end{equation}
It gives the transformation rules 
\begin{eqnarray*}
\Phi ^{\hspace{0.01in}\prime \prime } &=&\ul{\Phi }+\Phi _{0}^{\prime
},\qquad \breve{K}_{\phi }=K_{\phi }^{\prime }+\ul{K}_{\phi 0},\qquad 
\breve{K}_{C}=K_{C}^{\prime }+\ul{K}_{C0}, \\
\breve{K}_{\bar{C}}^{I} &=&K_{\bar{C}}^{I\hspace{0.01in}\prime
}-G^{Ii}(0,\partial )\phi ^{i\hspace{0.01in}\prime \prime }+\frac{\delta }{%
\delta \bar{C}^{I}}\int \mathcal{R}_{\bar{C}}^{J}(\bar{C},\ul{C}%
)K_{B}^{J\hspace{0.01in}\prime },\qquad \breve{K}_{B}=K_{B}^{\prime },
\end{eqnarray*}
which turn formula (\ref{uio})\ into 
\begin{eqnarray}
\hat{S}_{R}^{\prime }(\Phi ^{\prime \prime },K^{\prime }) &=&\hat{S}_{R}(0,%
\ul{\phi },0)-\int R_{\phi }^{i}(\ul{\Phi })\breve{K}_{\phi
}^{i}-\int R_{C}^{I}(\ul{\Phi })\breve{K}_{C}^{I}  \nonumber \\
&&-\int B^{I}\breve{K}_{\bar{C}}^{I}-\int \mathcal{R}_{\bar{C}}^{I}(\bar{C},%
\ul{C})\breve{K}_{\bar{C}}^{I}-\int \mathcal{R}_{B}^{I}(B,\ul{C%
})\breve{K}_{B}^{I}.  \label{fin0}
\end{eqnarray}

Note that $(\hat{S}_{R}^{\prime },\hat{S}_{R}^{\prime })=0$ is automatically
satisfied by (\ref{fin0}). Indeed, we know that $\hat{S}_{R}(\Phi ,%
\ul{\phi },K)$ is invariant under background transformations, and so
is $\hat{S}_{R}(0,\ul{\phi },0)$, because $\Phi $ and $K$ transform
as matter fields.

We can classify $\hat{S}_{R}(0,\ul{\phi },0)$ using its gauge
invariance. Let $\mathcal{G}_{i}(\phi )$ denote a basis of gauge-invariant
local functionals constructed with the physical fields $\phi $. Then 
\begin{equation}
\hat{S}_{R}(0,\ul{\phi },0)=\sum_{i}\tau _{i}\mathcal{G}_{i}(%
\ul{\phi }),  \label{comple}
\end{equation}
for suitable constants $\tau _{i}$.

Now we manipulate these results in several ways to make their consequences
more explicit. To prepare the next discussion it is convenient to relabel $\{%
\ul{\phi },\ul{C},\bar{C},B\}$ as $\breve{\Phi}^{\alpha }$ and 
$K^{\prime }$ as $K^{\prime \prime }$. Then formulas (\ref{for}) and (\ref
{fin0}) tell us that the canonical transformation 
\begin{equation}
F_{1}(\breve{\Phi},K^{\prime \prime })=\int \breve{\phi}K_{\phi }^{\prime
\prime }+\int \ \breve{C}K_{C}^{\prime \prime }+F_{R}(\{0,0,\brevebar{C},%
\breve{B}\},\{\breve{\phi},\breve{C}\},K^{\prime \prime },0,1)
\label{forex}
\end{equation}
is such that 
\begin{equation}
\hat{S}_{R}^{\prime }(\Phi ^{\prime \prime },K^{\prime \prime })=\hat{S}%
_{R}(0,\breve{\phi},0)-\int \bar{R}^{\alpha }(\breve{\Phi})\breve{K}_{\alpha
}.  \label{fin0ex}
\end{equation}

\paragraph{Parametric completeness\newline
}

Making the further canonical transformation $\breve{\Phi},\breve{K}%
\rightarrow \Phi ,K$ generated by 
\[
F_{2}(\Phi ,\breve{K})=\int \Phi ^{\alpha }\breve{K}_{\alpha }+\int \bar{C}%
^{I}G^{Ii}(0,\partial )\phi ^{i}-\int \mathcal{R}_{\bar{C}}^{I}(\bar{C},C)%
\breve{K}_{B}^{I}, 
\]
we get 
\begin{equation}
\hat{S}_{R}^{\prime }(\Phi ^{\prime \prime },K^{\prime \prime })=\hat{S}%
_{R}(0,\phi ,0)-\int R^{\alpha }(\Phi )\bar{K}_{\alpha },  \label{keynb}
\end{equation}
where the barred sources are the ones of (\ref{chif2}) at $\ul{\phi }%
=0$. If we start from the most general gauge-invariant classical action, 
\begin{equation}
S_{c}(\phi ,\lambda )\equiv \sum_{i}\lambda _{i}\mathcal{G}_{i}(\phi ),
\label{scgen}
\end{equation}
where $\lambda _{i}$ are physical couplings (apart from normalization constants),
identities (\ref{comple}) and (\ref{keynb}) give 
\begin{equation}
\hat{S}_{R}^{\prime }(\Phi ^{\prime \prime },K^{\prime \prime })=S_{c}(\phi
,\tau (\lambda ))-\int R^{\alpha }(\Phi )\bar{K}_{\alpha }.  \label{kk}
\end{equation}
This result proves parametric completeness in the usual approach, because it
tells us that the renormalized action of the usual approach is equal to the
classical action $\hat{S}^{\prime }(\Phi ,K)$ (check (\ref{sr2c})-(\ref
{sr2gf}) at $\ul{\Phi }=\ul{K}=0$), apart from parameter
redefinitions $\lambda \rightarrow \tau $ and a canonical transformation. In
this derivation the role of the background field method is just to provide
the key tool to prove the statement.

We can also describe parametric completeness in the background field
approach. Making the canonical transformation $\breve{\Phi},\breve{K}%
\rightarrow \hat{\Phi},\hat{K}$ generated by 
\begin{equation}
F_{2}^{\prime }(\hat{\Phi},\breve{K})=\int \hat{\Phi}^{\alpha }\breve{K}%
_{\alpha }+\int \ul{\phi }^{i}\breve{K}_{\phi }^{i}+\int \hatbar{C}%
^{I}G^{Ii}(\ul{\phi },\partial )\hat{\phi}^{i}-\int \mathcal{R}_{\bar{%
C}}^{I}(\hatbar{C},\hat{C})\breve{K}_{B}^{I},  \label{cancan}
\end{equation}
formula (\ref{fin0ex}) becomes 
\begin{equation}
\hat{S}_{R}^{\prime }(\Phi ^{\prime \prime },K^{\prime \prime })=S_{c}(\hat{%
\phi}+\ul{\phi },\tau )-\int R^{\alpha }(\hat{\phi}+\ul{\phi },%
\hat{C},\hatbar{C},\hat{B})\widetilde{\hat{K}}_{\alpha },  \label{y0}
\end{equation}
where the relations between tilde and nontilde sources are the hat
versions of (\ref{chif}). Next, we make a $\ul{\Phi }$ translation on
the left-hand side of (\ref{y0}) applying the canonical transformation $\Phi
^{\prime \prime },K^{\prime \prime }$ $\rightarrow \Phi ^{\prime },K^{\prime
}$ generated by 
\begin{equation}
F_{3}(\Phi ^{\prime },K^{\prime \prime })=\int (\Phi ^{\alpha \hspace{0.01in}%
\prime }+\ul{\Phi }^{\alpha })K_{\alpha }^{\prime \prime }.
\label{traslacan}
\end{equation}
Doing so, $\hat{S}_{R}^{\prime }(\Phi ^{\prime \prime },K^{\prime \prime })$
is turned into $\hat{S}_{R}^{\prime }(\Phi ^{\prime }+\ul{\Phi }%
,K^{\prime })$.

At this point, we want to compare the result we have obtained with (\/\ref
{ei}). Recall that formula (\ref{ei}) involves the canonical transformation (%
\ref{fr}) at $\xi =1$. If we set $\ul{K}^{\prime }=0$ we project that
canonical transformation onto a canonical transformation $\Phi ,K\rightarrow
\Phi ^{\prime },K^{\prime }$ generated by $F_{R}(\Phi ,\ul{\Phi }%
,K^{\prime },0,1)$, where $\ul{\Phi }$ is regarded as a spectator.
Furthermore, it is convenient to set $\ul{C}=0$, because then formula
(\ref{ei}) turns into 
\[
\hat{S}_{R}^{\prime }(\Phi ^{\prime }+\ul{\Phi },K^{\prime })=\hat{S}%
_{R}(\Phi ,\ul{\phi },K), 
\]
where now primed fields and sources are related to the unprimed ones by the
canonical transformation generated by $F_{R}(\Phi ,\{\ul{\phi }%
,0\},K^{\prime },0,1)$. Finally, recalling that $\hat{S}_{R}^{\prime }(\Phi
^{\prime \prime },K^{\prime \prime })=\hat{S}_{R}^{\prime }(\Phi ^{\prime }+%
\ul{\Phi },K^{\prime })$ and using (\ref{y0}) we get the key formula
we wanted, namely 
\begin{equation}
\hat{S}_{R}(\Phi ,\ul{\phi },K)=S_{c}(\hat{\phi}+\ul{\phi }%
,\tau (\lambda ))-\int R^{\alpha }(\hat{\phi}+\ul{\phi },\hat{C},%
\hatbar{C},\hat{B})\widetilde{\hat{K}}_{\alpha }.  \label{key}
\end{equation}
Observe that formula (\ref{key0}) of the introduction is formula (\ref{key})
with antighosts and Lagrange multipliers switched off. Checking (\ref
{sbaccagf}), formula (\ref{key}) tells us that the renormalized
background field action $\hat{S}_{R}$ is equal to the classical
background field action $\hat{S}_{\text{gf}}$ up to parameter redefinitions $%
\lambda \rightarrow \tau $ and a canonical transformation. This proves
parametric completeness in the background field approach.

The canonical transformation $\Phi ,K\rightarrow \hat{\Phi},\hat{K}$
involved in formula (\ref{key}) is generated by the functional $\hat{F}%
_{R}(\Phi ,\ul{\phi },\hat{K})$ obtained composing the
transformations generated by $F_{R}(\Phi ,\{\ul{\phi },0\},K^{\prime
},0,1)$, $F_{1}(\breve{\Phi},K^{\prime \prime })$, $F_{2}^{\prime }(\hat{\Phi%
},\breve{K})$ and $F_{3}(\Phi ^{\prime },K^{\prime \prime })$ of formulas (%
\ref{fr}), (\ref{forex}), (\ref{cancan}) and (\ref{traslacan}) (at $%
\ul{C}=0$). Working out the composition it is easy to prove that 
\[
\hatbar{C}=\bar{C},\qquad \hat{B}=B,\qquad \hat{K}_{B}=K_{B},\qquad \hat{K}_{%
\bar{C}}^{I}-G^{Ii}(\ul{\phi },\partial )\hat{\phi}^{i}=K_{\bar{C}%
}^{I}-G^{Ii}(\ul{\phi },\partial )\phi ^{i}, 
\]
and therefore $\hat{F}_{R}(\Phi ,\ul{\phi },\hat{K})$ has the form 
\begin{equation}
\hat{F}_{R}(\Phi ,\ul{\phi },\hat{K})=\int \Phi ^{\alpha }\hat{K}%
_{\alpha }+\Delta \hat{F}(\phi ,C,\ul{\phi },\hat{K}_{\phi }^{i}-%
\hatbar{C}^{I}G^{Ii}(\ul{\phi },-\overleftarrow{\partial }),\hat{K}%
_{C}),  \label{frhat}
\end{equation}
where $\Delta \hat{F}=\mathcal{O}(\hbar )$-poles.

\section{Examples}

\setcounter{equation}{0}

In this section we give two examples, non-Abelian gauge field theories and
quantum gravity, which are also useful to familiarize oneself with the notation and
the tools used in the paper. We switch to Minkowski spacetime. The
dimensional-regularization technique is understood.

\subsection{Yang-Mills theory}

The first example is non-Abelian Yang-Mills theory with simple gauge group $%
G $ and structure constants $f^{abc}$, coupled to fermions $\psi ^{i}$ in
some representation described by anti-Hermitian matrices $T_{ij}^{a}$. The
classical action $S_{c}(\phi )$ can be restricted by power counting, or
enlarged to include all invariants of (\ref{scgen}). The nonminimal
non-gauge-fixed action $S$ is the sum $\hat{S}+\bar{S}$ of (\ref{deco}) and (%
\ref{sbar}). We find 
\begin{eqnarray*}
\hat{S} &=&S_{c}(\phi +\ul{\phi })+\int \hspace{0.01in}\left[ g(\bar{%
\psi}^{i}+\ul{\bar{\psi}}^{i})T_{ij}^{a}C^{a}K_{\psi }^{j}+g\bar{K}_{\psi
}^{i}T_{ij}^{a}C^{a}(\psi ^{j}+\ul{\psi }^{j})\right] \\
&&-\int \hspace{0.01in}\left[ (\ul{D}_{\mu }C^{a}+gf^{abc}A_{\mu
}^{b}C^{c})K^{\mu a}-\frac{1}{2}gf^{abc}C^{b}C^{c}K_{C}^{a}+B^{a}K_{\bar{C}%
}^{a}\right] ,
\end{eqnarray*}
and 
\begin{eqnarray}
\bar{S} &=&gf^{abc}\int \hspace{0.01in}\ul{C}^{b}(A_{\mu }^{c}K^{\mu
a}+C^{c}K_{C}^{a}+\bar{C}^{c}K_{\bar{C}}^{a}+B^{c}K_{B}^{a})+g\int \hspace{%
0.01in}\left[ \bar{\psi}^{i}T_{ij}^{a}\ul{C}^{a}K_{\psi }^{j}+\bar{K}_{\psi
}^{i}T_{ij}^{a}\ul{C}^{a}\psi ^{j}\right]  \nonumber \\
&&-\int \hspace{0.01in}\left[ (\ul{D}_{\mu }\ul{C}^{a})%
\ul{K}^{\mu a}-\frac{1}{2}gf^{abc}\ul{C}^{b}\ul{C}^{c}%
\ul{K}_{C}^{a}-g\ul{\bar{\psi}}^{i}T_{ij}^{a}\ul{C}^{a}%
\ul{K}_{\psi }^{j}-g\ul{\bar{K}}_{\psi }^{i}T_{ij}^{a}\ul{C}%
^{a}\ul{\psi }^{j}\right] .  \label{sbary}
\end{eqnarray}
The covariant derivative $\ul{D}_{\mu }$ is the background one; for
example $\ul{D}_{\mu }\Lambda ^{a}=\partial _{\mu }\Lambda
^{a}+gf^{abc}\ul{A}_{\mu }^{b}\Lambda ^{c}$. The first line of (\ref
{sbary}) shows that all quantum fields transform as matter fields under
background transformations. It is easy to check that $\hat{S}$ and $\bar{S}$
satisfy $\llbracket \hat{S},\hat{S}\rrbracket =\llbracket \hat{S},\bar{S}%
\rrbracket =\llbracket \bar{S},\bar{S}\rrbracket =0$.

A common background-preserving gauge fermion is 
\begin{equation}
\Psi =\int \bar{C}^{a}\left( -\frac{\lambda }{2}B^{a}+\ul{D}^{\mu
}A_{\mu }^{a}\right) ,  \label{seeym}
\end{equation}
and the gauge-fixed action $\hat{S}_{\text{gf}}=\hat{S}+\llbracket \hat{S}%
,\Psi \rrbracket $ reads 
\[
\hat{S}_{\text{gf}}=\hat{S}-\frac{\lambda }{2}\int (B^{a})^{2}+\int B^{a}%
\ul{D}^{\mu }A_{\mu }^{a}-\int \bar{C}^{a}\ul{D}^{\mu }(%
\ul{D}_{\mu }C^{a}+gf^{abc}A_{\mu }^{b}C^{c}). 
\]
Since the gauge fixing is linear in the quantum fields, the action $\hat{S}$
depends on the combination $K_{\mu }^{a}+\ul{D}_{\mu }\bar{C}^{a}$
and not on $K_{\mu }^{a}$ and $\bar{C}^{a}$ separately. From now on we
switch matter fields off, for simplicity, and set $\lambda =0$.

We describe renormalization using the approach of this paper. First we
concentrate on the standard power-counting renormalizable case, where 
\[
S_{c}(A,g)=-\frac{1}{4}\int F_{\mu \nu }^{a}(A,g)F^{\mu \nu \hspace{0.01in}%
a}(A,g),\qquad \qquad F_{\mu \nu }^{a}(A,g)=\partial _{\mu }A_{\nu
}^{a}-\partial _{\nu }A_{\mu }^{a}+gf^{abc}A_{\mu }^{b}A_{\nu }^{c}.
\]
The key formula (\ref{key}) gives 
\begin{eqnarray}
\hat{S}_{R}(\Phi ,\ul{A},K) &=&-\frac{Z}{4}\int F_{\mu \nu }^{a}(%
\hat{A}+\ul{A},g)F^{\mu \nu \hspace{0.01in}a}(\hat{A}+\ul{A}%
,g)+\int \hat{B}^{a}\ul{D}^{\mu }\hat{A}_{\mu }^{a}-\int \hat{B}^{a}%
\hat{K}_{\bar{C}}^{a}  \nonumber \\
&&+\int \hspace{0.01in}(\hat{K}^{\mu a}+\ul{D}^{\mu }\hatbar{C}^{a})(%
\ul{D}_{\mu }\hat{C}^{a}+gf^{abc}\hat{A}_{\mu }^{b}\hat{C}^{c})+\frac{%
1}{2}gf^{abc}\int \hat{C}^{b}\hat{C}^{c}\hat{K}_{C}^{a},  \label{sat}
\end{eqnarray}
where $Z$ is a renormalization constant. The most general canonical
transformation $\Phi ,K\rightarrow \hat{\Phi},\hat{K}$ that is compatible
with power counting, global gauge invariance and ghost number conservation
can be easily written down. Introducing unknown constants where necessary,
we find that its generating functional has the form 
\begin{eqnarray*}
\hat{F}_{R}(\Phi ,\ul{A},\hat{K}) &=&\int (Z_{A}^{1/2}A_{\mu }^{a}+%
\ul{Z}_{A}^{1/2}\ul{A}_{\mu }^{a})\hat{K}^{\mu a}+\int
Z_{C}^{1/2}C^{a}\hat{K}_{C}^{a}+\int Z_{\bar{C}}^{1/2}\bar{C}^{a}\hat{K}_{%
\bar{C}}^{a} \\
&&+\int (Z_{B}^{1/2}B^{a}+\alpha \ul{D}^{\mu }A_{\mu }^{a}+\beta
\partial ^{\mu }\ul{A}_{\mu }^{a}+\gamma gf^{abc}\ul{A}^{\mu
b}A_{\mu }^{c}+\delta gf^{abc}\bar{C}^{b}C^{c})\hat{K}_{B}^{a} \\
&&+\int \bar{C}^{a}(\zeta B^{a}+\xi \ul{D}^{\mu }A_{\mu }^{a}+\eta
\partial ^{\mu }\ul{A}_{\mu }^{a}+\theta gf^{abc}\ul{A}^{\mu
b}A_{\mu }^{c}+\chi gf^{abc}\bar{C}^{b}C^{c}) \\
&&+\int \sigma \hat{K}_{\bar{C}}^{a}\hat{K}_{B}^{a}+\int \tau gf^{abc}C^{a}%
\hat{K}_{B}^{b}\hat{K}_{B}^{c}.
\end{eqnarray*}
Inserting it in (\ref{sat}) and using the nonrenormalization of the $B$
and $K_{\bar{C}}$-dependent terms, we find $a=\beta =\gamma =\delta =\zeta
=\theta =\chi =\sigma =\tau =0$ and 
\begin{equation}
\xi =1-Z_{\bar{C}}^{1/2}Z_{A}^{1/2},\qquad \eta =-Z_{\bar{C}}^{1/2}%
\ul{Z}_{A}^{1/2},\qquad Z_{B}=Z_{\bar{C}}.  \label{zeta}
\end{equation}
It is easy to check that $Z_{\bar{C}}$ disappears from the right-hand side
of (\ref{sat}), so we can set $Z_{\bar{C}}=1$. Furthermore, we know that $%
\hat{S}_{R}(\Phi ,\ul{A},K)$ is invariant under background
transformations ($\llbracket \hat{S}_{R},\bar{S}\rrbracket =0$), which
requires $\ul{Z}_{A}=0$. Finally, the canonical transformation just
reads 
\[
\hat{F}_{R}(\Phi ,\ul{A},\hat{K})=\int Z_{A}^{1/2}A_{\mu }^{a}\hat{K}%
^{\mu a}+\int Z_{C}^{1/2}C^{a}\hat{K}_{C}^{a}+\int \bar{C}^{a}\hat{K}_{\bar{C%
}}^{a}+\int B^{a}\hat{K}_{B}^{a}+(1-Z_{A}^{1/2})\int \bar{C}^{a}(\ul{D%
}^{\mu }A_{\mu }^{a}),
\]
which contains the right number of independent renormalization constants and
is of the form (\ref{frhat}).

Defining $Z_{g}=Z^{-1/2}$ and $Z_{A}^{\prime }=ZZ_{A}$ we can describe
renormalization in a more standard way. Writing 
\[
\hat{S}_{R}(0,\hat{A}+\ul{A},0)=-\frac{1}{4}\int F_{\mu \nu
}^{a}(Z_{A}^{\prime \hspace{0.01in}1/2}A+Z_{g}^{-1}\ul{A}%
,gZ_{g})F^{\mu \nu \hspace{0.01in}a}(Z_{A}^{\prime \hspace{0.01in}%
1/2}A+Z_{g}^{-1}\ul{A},gZ_{g}),
\]
we see that $Z_{g}$ is the usual gauge-coupling renormalization constant,
while $Z_{A}^{\prime }$ and $Z_{g}^{-2}$ are the wave-function
renormalization constants of the quantum gauge field $A$ and the background
gauge field $\ul{A}$, respectively. We remark that the local
divergent canonical transformation $\Phi ,K\rightarrow \hat{\Phi},\hat{K}$
corresponds to a highly nontrivial, convergent but non-local canonical
transformation at the level of the $\Gamma $ functional. 

If the theory is not power-counting renormalizable, then we need to consider
the most general classical action, equal to the right-hand side of (\ref
{scgen}). Counterterms include vertices with arbitrary numbers of external $%
\Phi $ and $K$ legs. Nevertheless, the key formula (\ref{key}) ensures that
the renormalized action $\hat{S}_{R}$ remains exactly the same, up to
parameter redefinitions and a canonical transformation. The only difference
is that now even the canonical transformation $\hat{F}_{R}(\Phi ,\ul{%
\phi },\hat{K})$ of (\ref{frhat})\ becomes nonpolynomial and highly
nontrivial.

\subsection{Quantum gravity}

Having written detailed formulas for Yang-Mills theory, in the case of
quantum gravity we can just outline the key ingredients. In particular, we
stress that the linearity assumption is satisfied both in the first-order
and second-order formalisms, both using the metric $g_{\mu \nu }$ and the
vielbein $e_{\mu }^{a}$. For example, using the second-order formalism and
the vielbein, the symmetry transformations are encoded in the expressions 
\begin{eqnarray*}
-\int R^{\alpha }(\Phi )K_{\alpha } &=&\int (e_{\rho }^{a}\partial _{\mu
}C^{\rho }+C^{\rho }\partial _{\rho }e_{\mu }^{a}+C^{ab}e_{\mu b})K_{a}^{\mu
}+\int C^{\rho }(\partial _{\rho }C^{\mu })K_{\mu }^{C} \\
&&+\int (C^{ac}\eta _{cd}C^{db}+C^{\rho }\partial _{\rho
}C^{ab})K_{ab}^{C}-\int B_{\mu }K_{\bar{C}}^{\mu }-\int B_{ab}K_{\bar{C}%
}^{ab}, \\
-\int \bar{R}^{\alpha }(\Phi )K_{\alpha } &=&-\int R^{\alpha }(\Phi
)K_{\alpha }-\int (\bar{C}_{\rho }\partial _{\mu }C^{\rho }-C^{\rho
}\partial _{\rho }\bar{C}_{\mu })K_{\bar{C}}^{\mu }+\int \left( B_{\rho
}\partial _{\mu }C^{\rho }+C^{\rho }\partial _{\rho }B_{\mu }\right)
K_{B}^{\mu },
\end{eqnarray*}
in the minimal and nonminimal cases, respectively, where $C^{\mu }$ are the
ghosts of diffeomorphisms, $C^{ab}$ are the Lorentz ghosts and $\eta _{ab}$
is the flat-space metric. We see that both $R^{\alpha }(\Phi )$ and $\bar{R}%
^{\alpha }(\Phi )$ are at most quadratic in $\Phi $. Matter fields are also
fine, since vectors $A_{\mu }$, fermions $\psi $ and scalars $\varphi $
contribute with 
\begin{eqnarray*}
&&-\int (\partial _{\mu }C^{a}+gf^{abc}A_{\mu }^{b}C^{c}-C^{\rho }\partial
_{\rho }A_{\mu }^{a}-A_{\rho }^{a}\partial _{\mu }C^{\rho })K_{A}^{\mu
a}+\int \left( C^{\rho }\partial _{\mu }C^{a}+\frac{1}{2}gf^{abc}C^{b}C^{c}%
\right) K_{C}^{a} \\
&&\qquad +\int C^{\rho }(\partial _{\rho }\varphi )K_{\varphi }+\int C^{\rho
}(\partial _{\rho }\bar{\psi})K_{\psi }-\frac{i}{4}\int \bar{\psi}\sigma
^{ab}C_{ab}K_{\psi }+\int K_{\bar{\psi}}C^{\rho }(\partial _{\rho }\psi )-%
\frac{i}{4}\int K_{\bar{\psi}}\sigma ^{ab}C_{ab}\psi ,
\end{eqnarray*}
where $\sigma ^{ab}=i[\gamma ^{a},\gamma ^{b}]/2$. Expanding around flat
space, common linear gauge-fixing conditions for diffeomorphisms and local Lorentz
symmetry are $\eta ^{\mu \nu }\partial _{\mu }e_{\nu }^{a}=\xi \eta ^{a\mu
}\partial _{\mu }e_{\nu }^{b}\delta _{b}^{\nu }$, $e_{\mu }^{a}=e_{\nu
}^{b}\eta _{b\mu }\eta ^{\nu a}$, respectively.

In the first-order formalism we just need to add the transformation of the
spin connection $\omega _{\mu }^{ab}$, encoded in 
\[
\int (C^{\rho }\partial _{\rho }\omega _{\mu }^{ab}+\omega _{\rho
}^{ab}\partial _{\mu }C^{\rho }-\partial _{\mu }C^{ab}+C^{ac}\eta
_{cd}\omega _{\mu }^{db}-\omega _{\mu }^{ac}\eta _{cd}C^{db})K_{ab}^{\mu }. 
\]
Moreover, in this case we can also gauge-fix local Lorentz symmetry with the
linear gauge-fixing condition $\eta ^{\mu \nu }\partial _{\mu }\omega _{\nu }^{ab}=0$,
instead of $e_{\mu }^{a}=e_{\nu }^{b}\eta _{b\mu }\eta ^{\nu a}$.

We see that all gauge symmetries that are known to have physical interest
satisfy the linearity assumption, together with irreducibility and off-shell
closure. On the other hand, more speculative symmetries (such as local
supersymmetry) do not satisfy those assumptions in an obvious way. When
auxiliary fields are introduced to achieve off-shell closure, some symmetry
transformations (typically, those of auxiliary fields) are nonlinear \cite
{superg}. The relevance of this issue is already known in the literature.
For example, in ref. \cite{superspace} it is explained that in
supersymmetric theories the standard background field method cannot be
applied, precisely because the symmetry transformations are nonlinear. It is
argued that the linearity assumption is tied to the linear splitting $\Phi
\rightarrow \Phi +\ul{\Phi }$ between quantum fields $\Phi $ and
background fields $\ul{\Phi }$, and that the problem of
supersymmetric theories can be avoided with a nonlinear splitting of the
form $\Phi \rightarrow \Phi +\ul{\Phi }$ $+$ nonlinear corrections.
Perhaps it is possible to generalize the results of this paper to
supergravity following those guidelines. From our viewpoint, the crucial
property is that the background transformations of the quantum fields are
linear in the quantum fields themselves, because then they do not
renormalize.

An alternative strategy, not bound to supersymmetric theories, is that of
introducing (possibly infinitely many) auxiliary fields, replacing\ every
nonlinear term appearing in the symmetry transformations with a new field $%
N $, and then proceeding similarly with the $N$ transformations and the
closure relations, till all functions $R^{\alpha }$ are at most quadratic.
The natural framework for this kind of job is the one of refs. \cite
{fieldcov,masterf,mastercan}, where the fields $N$ are dual to the sources $L$ coupled to
composite fields. Using that approach the Batalin-Vilkovisky formalism can
be extended to the composite-field sector of the theory and all perturbative
canonical transformations can be studied as true changes of field variables
in the functional integral, instead of mere replacements of integrands. For
reasons of space, though, we cannot pursue this strategy here.

\section{The quest for parametric completeness: Where we stand now}

\setcounter{equation}{0}

In this section we make remarks about the problem of parametric completeness in general gauge theories
and recapitulate where we stand now on this issue. To begin with, consider
non-Abelian Yang-Mills theory as a deformation of its Abelian limit. The
minimal solution $S(g)$ of the master equation $(S(g),S(g))=0$ reads 
\[
S(g)=-\frac{1}{4}\int F_{\mu \nu }^{a}F^{\mu \nu \hspace{0.01in}a}+\int
K^{\mu a}\partial _{\mu }C^{a}+gf^{abc}\int \left( K^{\mu a}A_{\mu
}^{b}+\frac{1}{2}K_{C}^{a}C^{b}\right) C^{c}.
\]
Differentiating the master equation with respect to $g$ and setting $g=0$,
we find 
\[
(S,S)=0,\qquad (S,\omega )=0,\qquad S=S(0),\qquad \omega =\left. \frac{%
\mathrm{d}S(g)}{\mathrm{d}g}\right| _{g=0}.
\]
On the other hand, we can easily prove that there exists no local functional 
$\chi $ such that $\omega =(S,\chi )$. Thus, we can say that $\omega $ is a
nontrivial solution of the cohomological problem associated with an Abelian
Yang-Mills theory that contains a suitable number of photons \cite{regnocoho}%
. Nevertheless, renormalization cannot turn $\omega $ on as a counterterm,
because $S(0)$ is a free field theory. Even if we couple the theory to
gravity and assume that massive fermions are present (which allows us to
construct dimensionless parameters multiplying masses with the Newton
constant), radiative corrections cannot dynamically ``un-Abelian-ize'' the
theory, namely convert an Abelian theory into a non-Abelian one. One way to
prove this fact is to note that the dependence on gauge fields is even at $%
g=0$, but not at $g\neq 0$. The point is, however, that cohomology \textit{%
per se} is unable to prove it. Other properties must be advocated, such
as the discrete symmetry just mentioned. In general, we cannot rely on
cohomology only, and the possibility that gauge symmetries may be
dynamically deformed in nontrivial and observable ways remains open.

In ref. \cite{regnocoho} the issue of parametric completeness was studied in
general terms. In that approach, which applies to all theories that are
manifestly free of gauge anomalies, renormalization triggers an automatic
parametric extension till the classical action becomes parametrically
complete. The results of ref. \cite{regnocoho} leave the door open to
dynamically induced nontrivial deformations of the gauge symmetry. Instead,
the results found here close that door in all cases where they apply, which
means manifestly nonanomalous irreducible gauge symmetries that close
off shell and satisfy the linearity assumption. The reason is -- we stress
it again -- that by formulas (\ref{kk}) and (\ref{key}) all dynamically
induced deformations can be organized into parameter redefinitions and
canonical transformations. As far as we know now, gauge symmetries can still
be dynamically deformed in observable ways in theories that do not satisfy
the assumptions of this paper. Supergravities are natural candidates to
provide explicit examples. 

\section{Conclusions}

\setcounter{equation}{0}

The background field method and the Batalin-Vilkovisky formalism are
convenient tools to quantize general gauge field theories. In this paper we have merged the two to rephrase and generalize known results about
renormalization, and to study parametric completeness. Our approach applies
when gauge anomalies are manifestly absent, the gauge algebra is irreducible
and closes off shell, the gauge transformations are linear functions of the
fields, and closure is field independent. These assumptions are sufficient
to include the gauge symmetries we need for physical applications, such as
Abelian and non-Abelian Yang-Mills symmetries, local Lorentz symmetry and
general changes of coordinates, but exclude other potentially interesting
symmetries, such as local supersymmetry. Both renormalizable and
nonrenormalizable theories are covered, such as QED, non-Abelian
Yang-Mills theories, quantum gravity and Lorentz-violating gauge theories, as
well as effective and higher-derivative models, in arbitrary dimensions, and
also extensions obtained adding any set of composite fields. At the same
time, chiral theories, and therefore the Standard Model, possibly coupled with
quantum gravity, require the analysis of anomaly cancellation and the
Adler-Bardeen theorem, which we postpone to a future investigation. The fact
that supergravities are left out from the start, on the other hand, suggests
that there should exist either a no-go theorem or a more advanced framework.
At any rate, we are convinced that our formalism is helpful to understand
several properties better and address unsolved problems.

We have studied gauge dependence in detail, and renormalized the canonical
transformation that continuously interpolates between the background field
approach and the usual approach. Relating the two approaches, we have proved
parametric completeness without making use of cohomological classifications.
The outcome is that in all theories that satisfy our assumptions
renormalization cannot hide any surprises; namely the gauge symmetry remains
essentially the same throughout the quantization process. In the theories that
do not satisfy our assumptions, instead, the gauge symmetry could be
dynamically deformed in physically observable ways. It would be remarkable
if we discovered explicit examples of theories where this sort of
``dynamical creation'' of gauge symmetries actually takes place.

\vskip 12truept \noindent {\large \textbf{Acknowledgments}}

\vskip 2truept

The investigation of this paper was carried out as part of a program to
complete the book \cite{webbook}, which will be available at %
\href{http://renormalization.com}{\texttt{Renormalization.com}} once
completed.

\renewcommand{\thesection}{A}

\section{Appendix}

\renewcommand{\theequation}{\thesection.\arabic{equation}}

In this appendix we prove several theorems and identities that are used in
the paper. We use the Euclidean notation and the dimensional-regularization
technique, which guarantees, in particular, that the functional integration
measure is invariant under perturbatively local changes of field variables.
The generating functionals $Z$ and $W$ are defined from 
\begin{equation}
Z(J,K,\ul{\Phi },\ul{K})=\int [\mathrm{d}\Phi ]\hspace{0.01in}%
\exp (-S(\Phi ,\ul{\Phi },K,\ul{K})+\int \Phi ^{\alpha
}J_{\alpha })=\exp W(J,K,\ul{\Phi },\ul{K}),  \label{defa}
\end{equation}
and $\Gamma (\Phi ,\ul{\Phi },K,\ul{K})$ $=-W+\int \Phi
^{\alpha }J_{\alpha }$ is the $W$ Legendre transform. Averages denote the
sums of connected diagrams (e.g. $\langle A(x)B(y)\rangle =\langle
A(x)B(y)\rangle _{\text{nc}}-\langle A(x)\rangle \langle B(y)\rangle $,
where $\langle A(x)B(y)\rangle _{\text{nc}}$ includes disconnected
diagrams). Moreover, the average $\langle X\rangle $ of a local functional $%
X $ can be viewed as a functional of $\Phi ,\ul{\Phi },K,\ul{K}
$ (in which case it collects one-particle irreducible diagrams) or a
functional of $J,\ul{\Phi },K,\ul{K}$. When we need to
distinguish the two options we specify whether $\Phi $ or $J$ are kept
constant in functional derivatives. First we work in the usual
(non-background field) framework; then we generalize the results to the
background field method.

To begin with, we recall a property that is true even when the action $S(\Phi ,K)$ does not
satisfy the master equation.

\begin{theorem}
The identity $(\Gamma ,\Gamma )=\langle (S,S)\rangle $ holds. \label{th0}
\end{theorem}

\textit{Proof}. Applying the change of field variables 
\begin{equation}
\Phi ^{\alpha }\rightarrow \Phi ^{\alpha }+\theta (S,\Phi ^{\alpha })
\label{chv}
\end{equation}
to (\ref{defa}), where $\theta $ is a constant anticommuting parameter, we
obtain 
\[
\theta \int \left\langle \frac{\delta _{r}S}{\delta K_{\alpha }}\frac{\delta
_{l}S}{\delta \Phi ^{\alpha }}\right\rangle -\theta \int \left\langle \frac{%
\delta _{r}S}{\delta K_{\alpha }}\right\rangle J_{\alpha }=0, 
\]
whence 
\[
\frac{1}{2}\langle (S,S)\rangle =-\int \left\langle \frac{\delta _{r}S}{%
\delta K_{\alpha }}\frac{\delta _{l}S}{\delta \Phi ^{\alpha }}\right\rangle
=-\int \left\langle \frac{\delta _{r}S}{\delta K_{\alpha }}\right\rangle
J_{\alpha }=\int \frac{\delta _{r}W}{\delta K_{\alpha }}\frac{\delta
_{l}\Gamma }{\delta \Phi ^{\alpha }}=-\int \frac{\delta _{r}\Gamma }{\delta
K_{\alpha }}\frac{\delta _{l}\Gamma }{\delta \Phi ^{\alpha }}=\frac{1}{2}%
(\Gamma ,\Gamma ). 
\]

Now we prove results for an action $S$ that satisfies the master equation $%
(S,S)=0$.

\begin{theorem}
If $(S,S)=0$ then $(\Gamma ,\langle X\rangle )=\langle (S,X)\rangle $ for
every local functional $X$. \label{theorem2}
\end{theorem}

\textit{Proof}. Applying the change of field variables (\ref{chv}) to 
\[
\langle X\rangle =\frac{1}{Z(J,K)}\int [\mathrm{d}\Phi ]\hspace{0.01in}X\exp
(-S+\int \Phi ^{\alpha }J_{\alpha }), 
\]
and using $(S,S)=0$ we obtain 
\begin{equation}
\int \left\langle \frac{\delta _{r}S}{\delta K_{\alpha }}\frac{\delta _{l}X}{%
\delta \Phi ^{\alpha }}\right\rangle =(-1)^{\varepsilon _{X}+1}\int
\left\langle X\frac{\delta _{r}S}{\delta K_{\alpha }}\right\rangle \frac{%
\delta _{l}\Gamma }{\delta \Phi ^{\alpha }},  \label{r1}
\end{equation}
where $\varepsilon _{X}$ denotes the statistics of the functional $X$ (equal
to 0 if $X$ is bosonic, 1 if it is fermionic, modulo 2). Moreover, we also
have 
\begin{equation}
\int \left\langle \frac{\delta _{r}S}{\delta \Phi ^{\alpha }}\frac{\delta
_{l}X}{\delta K_{\alpha }}\right\rangle =\int \frac{\delta _{r}\Gamma }{%
\delta \Phi ^{\alpha }}\left\langle \frac{\delta _{l}X}{\delta K_{\alpha }}%
\right\rangle ,  \label{r2}
\end{equation}
which can be proved starting from the expression on the left-hand side and
integrating by parts. In the derivation we use the fact that since $X$ is
local, $\delta _{r}\delta _{l}X/(\delta \Phi ^{\alpha }\delta K_{\alpha })$
is set to zero by the dimensional regularization, which kills the $\delta
(0) $s and their derivatives.

Next, straightforward differentiations give 
\begin{eqnarray}
\left. \frac{\delta _{l}\langle X\rangle }{\delta K_{\alpha }}\right| _{J}
&=&\left\langle \frac{\delta _{l}X}{\delta K_{\alpha }}\right\rangle
-\left\langle \frac{\delta _{l}S}{\delta K_{\alpha }}X\right\rangle
\label{r3} \\
&=&\left. \frac{\delta _{l}\langle X\rangle }{\delta K_{\alpha }}\right|
_{\Phi }-\int \left. \frac{\delta _{l}J_{\beta }}{\delta K_{\alpha }}\right|
_{\Phi }\left. \frac{\delta _{l}\langle X\rangle }{\delta J_{\beta }}\right|
_{K}.  \label{r4}
\end{eqnarray}
At this point, using (\ref{r1})-(\ref{r4}) and $(J_{\alpha},\Gamma )=0$ (which
can be proved by differentiating $(\Gamma ,\Gamma )=0$ with respect to $\Phi
^{\alpha}$), we derive $(\Gamma ,\langle X\rangle )=\langle (S,X)\rangle $.

\begin{corollary}
If $(S,S)=0$ and 
\begin{equation}
\frac{\partial S}{\partial \xi }=(S,X),  \label{bbug}
\end{equation}
where $X$ is a local functional and $\xi $ is a parameter, then 
\begin{equation}
\frac{\partial \Gamma }{\partial \xi }=(\Gamma ,\langle X\rangle ).
\label{pprove}
\end{equation}
\label{bbugc}
\end{corollary}

\textit{Proof}. Using theorem \ref{theorem2} we have 
\[
\frac{\partial \Gamma }{\partial \xi }=-\frac{\partial W}{\partial \xi }%
=\langle \frac{\partial S}{\partial \xi }\rangle =\langle (S,X)\rangle
=(\Gamma ,\langle X\rangle ). 
\]

\medskip

Now we derive results that hold even when $S$ does not satisfy the master
equation.

\begin{theorem}
\label{blabla}The identity 
\begin{equation}
(\Gamma ,\langle X\rangle )=\langle (S,X)\rangle -\frac{1}{2}\langle
(S,S)X\rangle _{\Gamma }  \label{prove0}
\end{equation}
holds, where $X$ is a generic local functional and $\langle AB\cdots
Z\rangle _{\Gamma }$ denotes the set of connected, one-particle irreducible
diagrams with one insertion of $A$, $B$, $\ldots Z$.
\end{theorem}

This theorem is a generalization of theorem \ref{theorem2}. It is proved
by repeating the derivation without using $\left( S,S\right) =0$. First,
observe that formula (\ref{r1}) generalizes to 
\begin{equation}
\int \left\langle \frac{\delta _{r}S}{\delta K_{\alpha }}\frac{\delta _{l}X}{%
\delta \Phi ^{\alpha }}\right\rangle =(-1)^{\varepsilon _{X}+1}\int
\left\langle X\frac{\delta _{r}S}{\delta K_{\alpha }}\right\rangle \frac{%
\delta _{l}\Gamma }{\delta \Phi ^{\alpha }}-\frac{1}{2}\langle (S,S)X\rangle
.  \label{r11}
\end{equation}
On the other hand, formula (\ref{r2}) remains the same, as well as (\ref{r3}%
) and (\ref{r4}). We have 
\[
\left( \Gamma ,\langle X\rangle \right) =\langle (S,X)\rangle -\frac{1}{2}%
\langle (S,S)X\rangle +\int \frac{\delta _{r}\Gamma }{\delta \Phi ^{\alpha }}%
\left. \frac{\delta _{l}J_{\beta }}{\delta K_{\alpha }}\right| _{\Phi
}\left. \frac{\delta _{l}\langle X\rangle }{\delta J_{\beta }}\right|
_{K}-\int \frac{\delta _{r}\Gamma }{\delta K_{\alpha }}\left. \frac{\delta
_{l}\langle X\rangle }{\delta \Phi ^{\alpha }}\right| _{K}. 
\]
Differentiating $(\Gamma ,\Gamma )$ with respect to $\Phi ^{\alpha}$ we get 
\[
\frac{1}{2}\frac{\delta _{r}(\Gamma ,\Gamma )}{\delta \Phi ^{\alpha}}=\frac{1}{2}%
\frac{\delta _{l}(\Gamma ,\Gamma )}{\delta \Phi ^{\alpha}}=(J_{\alpha },\Gamma
)=(-1)^{\varepsilon _{\alpha }}(\Gamma ,J_{\alpha }), 
\]
where $\varepsilon _{\alpha }$ is the statistics of $\Phi ^{\alpha }$. Using 
$(\Gamma ,\Gamma )=\langle (S,S)\rangle $ we finally obtain 
\begin{equation}
\left( \Gamma ,\langle X\rangle \right) =\langle (S,X)\rangle -\frac{1}{2}%
\langle (S,S)X\rangle +\frac{1}{2}\int (-1)^{\varepsilon _{\alpha }}\frac{%
\delta _{r}\langle (S,S)\rangle }{\delta \Phi ^{\alpha }}\left. \frac{\delta
_{l}\langle X\rangle }{\delta J_{\alpha }}\right| _{K}.  \label{allo}
\end{equation}

The set of irreducible diagrams contained in $\langle A\hspace{0.01in}%
B\rangle $, where $A$ and $B$ are local functionals, is given by the formula 
\begin{equation}
\langle A\hspace{0.01in}B\rangle _{\Gamma }=\langle AB\rangle -\{\langle
A\rangle ,\langle B\rangle \},  \label{oo}
\end{equation}
where $\{X,Y\}$ are the ``mixed brackets'' \cite{BV2} 
\begin{equation}
\{X,Y\}\equiv \int \frac{\delta _{r}X}{\delta \Phi ^{\alpha }}\langle \Phi
^{\alpha }\Phi ^{\beta }\rangle \frac{\delta _{l}Y}{\delta \Phi ^{\beta }}%
=\int \frac{\delta _{r}X}{\delta \Phi ^{\alpha }}\frac{\delta _{r}\delta
_{r}W}{\delta J_{\beta }\delta J_{\alpha }}\frac{\delta _{l}Y}{\delta \Phi
^{\beta }}=\int \left. \frac{\delta _{r}X}{\delta J_{\alpha }}\right| _{K}%
\frac{\delta _{l}Y}{\delta \Phi ^{\alpha }},  \label{mixed brackets}
\end{equation}
$X$ and $Y$ being functionals of $\Phi $ and $K$. Indeed, $\{\langle
A\rangle ,\langle B\rangle \}$ is precisely the set of diagrams
in which the $A$  and $B$ insertions are connected in a one-particle reducible way. Thus,
formula (\ref{allo}) coincides with (\ref{prove0}).

Using (\ref{prove0}) we also have the identity 
\begin{equation}
\frac{\partial \Gamma }{\partial \xi }-\left( \Gamma ,\langle X\rangle
\right) =\left\langle \frac{\partial S}{\partial \xi }-\left( S,X\right)
\right\rangle +\frac{1}{2}\left\langle \left( S,S\right) \hspace{0.01in}%
X\right\rangle _{\Gamma },  \label{provee}
\end{equation}
which generalizes corollary \ref{bbugc}.

\medskip

Now we switch to the background field method. We begin by generalizing theorem 
\ref{th0}.

\begin{theorem}
If the action $S(\Phi ,\ul{\Phi },K,\ul{K})$ is such that $%
\delta _{l}S/\delta \ul{K}_{\alpha }$ is $\Phi $ independent, the
identity 
\[
\llbracket \Gamma ,\Gamma \rrbracket =\langle \llbracket S,S\rrbracket %
\rangle 
\]
holds. \label{thb}
\end{theorem}

\textit{Proof}. Since $\delta _{l}S/\delta \ul{K}_{\alpha }$ is $\Phi 
$ independent we have $\delta _{l}\Gamma /\delta \ul{K}_{\alpha
}=\delta _{l}S/\delta \ul{K}_{\alpha }$. Using theorem \ref{th0} we
find 
\[
\llbracket \Gamma ,\Gamma \rrbracket =(\Gamma ,\Gamma )+2\int \frac{\delta
_{r}\Gamma }{\delta \ul{\Phi }^{\alpha }}\frac{\delta _{l}\Gamma }{%
\delta \ul{K}_{\alpha }}=\langle (S,S)\rangle +2\int \langle \frac{%
\delta _{r}S}{\delta \ul{\Phi }^{\alpha }}\rangle \frac{\delta _{l}S}{%
\delta \ul{K}_{\alpha }}=\langle (S,S)\rangle +2\int \langle \frac{%
\delta _{r}S}{\delta \ul{\Phi }^{\alpha }}\frac{\delta _{l}S}{\delta 
\ul{K}_{\alpha }}\rangle =\langle \llbracket S,S\rrbracket \rangle . 
\]

Next, we mention the useful identity 
\begin{equation}
\left. \frac{\delta _{l}\langle X\rangle }{\delta \ul{\Phi }^{\alpha }%
}\right| _{\Phi }=\left\langle \frac{\delta _{l}X}{\delta \ul{\Phi }%
^{\alpha }}\right\rangle -\left\langle \frac{\delta _{l}S}{\delta \ul{%
\Phi }^{\alpha }}X\right\rangle _{\Gamma },  \label{dera}
\end{equation}
which holds for every local functional $X$. It can be proved by taking (\ref
{r3})--(\ref{r4}) with $K\rightarrow \ul{\Phi }$ and using (\ref{oo})--(%
\ref{mixed brackets}).

Mimicking the proof of theorem \ref{thb} and using (\ref{dera}), it is easy
to prove that theorem \ref{blabla} implies the identity 
\begin{equation}
\llbracket \Gamma ,\langle X\rangle \rrbracket =\langle \llbracket S,X%
\rrbracket \rangle -\frac{1}{2}\langle \llbracket S,S\rrbracket X\rangle
_{\Gamma },  \label{bb2}
\end{equation}
for every $\ul{K}$-independent local functional $X$. Thus we have the
following property.

\begin{corollary}
The identity 
\begin{equation}
\frac{\partial \Gamma }{\partial \xi }-\llbracket \Gamma ,\langle X\rangle %
\rrbracket =\left\langle \frac{\partial S}{\partial \xi }-\llbracket S,X%
\rrbracket \right\rangle +\frac{1}{2}\langle \llbracket S,S\rrbracket %
X\rangle _{\Gamma }  \label{proveg}
\end{equation}
holds for every action $S$ such that $\delta _{l}S/\delta \ul{K}%
_{\alpha }$ is $\Phi $ independent, for every $\ul{K}$-independent local
functional $X$ and for every parameter $\xi $. \label{cora}
\end{corollary}

If the action $S$ has the form (\ref{assu}) and $X$ is also $\ul{C}$
independent, applying (\ref{backghost}) to (\ref{bb2}) we obtain 
\begin{equation}
\llbracket \hat{\Gamma},\langle X\rangle \rrbracket =\langle \llbracket \hat{%
S},X\rrbracket \rangle -\frac{1}{2}\langle \llbracket \hat{S},\hat{S}%
\rrbracket X\rangle _{\Gamma },\qquad \llbracket \bar{S},\langle X\rangle %
\rrbracket =\langle \llbracket \bar{S},X\rrbracket \rangle -\langle %
\llbracket \bar{S},\hat{S}\rrbracket X\rangle _{\Gamma },\qquad \langle %
\llbracket \bar{S},\bar{S}\rrbracket X\rangle _{\Gamma }=0,
\label{blablaback}
\end{equation}
which imply the following statement.

\begin{corollary}
If $S$ satisfies the assumptions of (\ref{assu}), $\llbracket \bar{S},X%
\rrbracket =0$ and $\llbracket \bar{S},\hat{S}\rrbracket =0$, where $X$ is a 
$\ul{C}$- and $\ul{K}$-independent local functional, then $%
\llbracket \bar{S},\langle X\rangle \rrbracket =0$. \label{corolla}
\end{corollary}

Finally, we recall a result derived in ref. \cite{removal}.

\begin{theorem}
If $\Phi ,K\rightarrow \Phi ^{\prime },K^{\prime }$ is a canonical
transformation generated by $F(\Phi ,K^{\prime })$, and $\chi (\Phi ,K)$ is
a functional behaving as a scalar (that is to say $\chi ^{\prime }(\Phi ^{\prime
},K^{\prime })=\chi (\Phi ,K)$), then 
\begin{equation}
\frac{\partial \chi ^{\prime }}{\partial \varsigma }=\frac{\partial \chi }{%
\partial \varsigma }-(\chi ,\tilde{F}_{\varsigma })  \label{thesis}
\end{equation}
for every parameter $\varsigma $, where $\tilde{F}_{\varsigma }(\Phi
,K)\equiv F_{\varsigma }(\Phi ,K^{\prime }(\Phi ,K))$ and $F_{\varsigma
}(\Phi ,K^{\prime })=\partial F/\partial \varsigma $. \label{theorem5}
\end{theorem}

\textit{Proof}. When we do not specify the variables that are kept constant
in partial derivatives, it is understood that they are the natural
variables. Thus $F$, $\Phi ^{\prime }$ and $K$ are functions of $\Phi
,K^{\prime }$, while $\chi $ and $\tilde{F}_{\varsigma }$ are functions of $%
\Phi ,K$ and $\chi ^{\prime }$ is a function of $\Phi ^{\prime },K^{\prime }$%
. It is useful to write down the differentials of $\Phi ^{\prime }$ and $K$,
which are \cite{vanproeyen} 
\begin{eqnarray}
\mathrm{d}\Phi ^{\prime \hspace{0.01in}\alpha } &=&\int \frac{\delta
_{l}\delta F}{\delta K_{\alpha }^{\prime }\delta \Phi ^{\beta }}\mathrm{d}%
\Phi ^{\beta }+\int \frac{\delta _{l}\delta F}{\delta K_{\alpha }^{\prime
}\delta K_{\beta }^{\prime }}\mathrm{d}K_{\beta }^{\prime }+\frac{\partial
\Phi ^{\prime \hspace{0.01in}\alpha }}{\partial \varsigma }\mathrm{d}%
\varsigma ,  \nonumber \\
\mathrm{d}K_{\alpha } &=&\int \mathrm{d}\Phi ^{\beta }\frac{\delta
_{l}\delta F}{\delta \Phi ^{\beta }\delta \Phi ^{\alpha }}+\int \mathrm{d}%
K_{\beta }^{\prime }\frac{\delta _{l}\delta F}{\delta K_{\beta }^{\prime
}\delta \Phi ^{\alpha }}+\frac{\partial K_{\alpha }}{\partial \varsigma }%
\mathrm{d}\varsigma .  \label{differentials}
\end{eqnarray}

Differentiating $\chi ^{\prime }(\Phi ^{\prime },K^{\prime })=\chi (\Phi ,K)$
with respect to $\varsigma $ at constant $\Phi ^{\prime }$ and $K^{\prime }$%
, we get 
\begin{equation}
\frac{\partial \chi ^{\prime }}{\partial \varsigma }=\frac{\partial \chi }{%
\partial \varsigma }+\int \frac{\delta _{r}\chi }{\delta \Phi ^{\alpha }}%
\left. \frac{\partial \Phi ^{\alpha }}{\partial \varsigma }\right| _{\Phi
^{\prime },K^{\prime }}+\int \frac{\delta _{r}\chi }{\delta K_{\alpha }}%
\left. \frac{\partial K_{\alpha }}{\partial \varsigma }\right| _{\Phi
^{\prime },K^{\prime }}.  \label{sigmaprimosue2}
\end{equation}
Formulas (\ref{differentials}) allow us to write 
\[
\frac{\partial \Phi ^{\prime \hspace{0.01in}\alpha }}{\partial \varsigma }%
=-\int \frac{\delta _{l}\delta F}{\delta K_{\alpha }^{\prime }\delta \Phi
^{\beta }}\left. \frac{\partial \Phi ^{\beta }}{\partial \varsigma }\right|
_{\Phi ^{\prime },K^{\prime }},\qquad \frac{\delta _{l}\delta F}{\delta
K_{\alpha }^{\prime }\delta \Phi ^{\beta }}=\left. \frac{\delta _{l}K_{\beta
}}{\delta K_{\alpha }^{\prime }}\right| _{\Phi ,\varsigma }, 
\]
and therefore we have 
\begin{equation}
\frac{\delta \tilde{F}_{\varsigma }}{\delta K_{\alpha }}=\int \left. \frac{%
\delta _{l}K_{\beta }^{\prime }}{\delta K_{\alpha }}\right| _{\Phi
,\varsigma }\frac{\partial \Phi ^{\prime \hspace{0.01in}\beta }}{\partial
\varsigma }=-\int \left. \frac{\delta _{l}K_{\beta }^{\prime }}{\delta
K_{\alpha }}\right| _{\Phi ,\varsigma }\left. \frac{\delta _{l}K_{\gamma }}{%
\delta K_{\beta }^{\prime }}\right| _{\Phi ,\varsigma }\left. \frac{\partial
\Phi ^{\gamma }}{\partial \varsigma }\right| _{\Phi ^{\prime },K^{\prime
}}=-\left. \frac{\partial \Phi ^{\alpha }}{\partial \varsigma }\right|
_{\Phi ^{\prime },K^{\prime }}.  \label{div1}
\end{equation}
Following analogous steps, we also find 
\[
\frac{\delta \tilde{F}_{\varsigma }}{\delta \Phi ^{\alpha }}=\frac{\partial
K_{\alpha }}{\partial \varsigma }+\int \left. \frac{\delta _{l}K_{\beta
}^{\prime }}{\delta \Phi ^{\alpha }}\right| _{K,\varsigma }\frac{\partial
\Phi ^{\prime \hspace{0.01in}\beta }}{\partial \varsigma },\qquad \frac{%
\partial K_{\alpha }}{\partial \varsigma }=\left. \frac{\partial K_{\alpha }%
}{\partial \varsigma }\right| _{\Phi ^{\prime },K^{\prime }}-\int \frac{%
\delta _{l}\delta F}{\delta \Phi ^{\alpha }\delta \Phi ^{\beta }}\left. 
\frac{\partial \Phi ^{\beta }}{\partial \varsigma }\right| _{\Phi ^{\prime
},K^{\prime }}, 
\]
whence 
\begin{equation}
\left. \frac{\partial K_{\alpha }}{\partial \varsigma }\right| _{\Phi
^{\prime },K^{\prime }}=\frac{\delta \tilde{F}_{\varsigma }}{\delta \Phi
^{\alpha }}+\int \left( \frac{\delta _{l}K_{\gamma }}{\delta \Phi ^{\alpha }}%
+\left. \frac{\delta _{l}K_{\beta }^{\prime }}{\delta \Phi ^{\alpha }}%
\right| _{K,\varsigma }\frac{\delta _{l}K_{\gamma }}{\delta K_{\beta
}^{\prime }}\right) \left. \frac{\partial \Phi ^{\gamma }}{\partial
\varsigma }\right| _{\Phi ^{\prime },K^{\prime }}=\frac{\delta \tilde{F}%
_{\varsigma }}{\delta \Phi ^{\alpha }}.  \label{div2}
\end{equation}
This formula, together with (\ref{div1}), allows us to rewrite (\ref
{sigmaprimosue2}) in the form (\ref{thesis}).


\begin{thebibliography}{99}
\bibitem{dewitt}  B.S. De Witt, Quantum theory of gravity. II. The
manifestly covariant theory, Phys. Rev. 162 (1967) 1195.

\bibitem{abbott}  L.F. Abbott, The background field method beyond one loop,
Nucl. Phys. B 185 (1981) 189.

\bibitem{bata}  I.A. Batalin and G.A. Vilkovisky, Gauge algebra and
quantization, Phys. Lett. B 102 (1981) 27-31;

\noindent I.A. Batalin and G.A. Vilkovisky, Quantization of gauge theories
with linearly dependent generators, Phys. Rev. D 28 (1983) 2567,
Erratum-ibid. D 30 (1984) 508;

see also S. Weinberg, \textit{The quantum theory of fields}, vol. II,
Cambridge University Press, Cambridge 1995.

\bibitem{regnocoho}  D. Anselmi, Renormalization of gauge theories without
cohomology, Eur. Phys. J. C73 (2013) 2508, %
\href{http://renormalization.com/13a1/}{13A1 Renormalization.com},
arXiv:1301.7577 [hep-th].

\bibitem{coho}  G. Barnich, F. Brandt, M. Henneaux, Local BRST cohomology in
the antifield formalism. I. General theorems, Commun. Math. Phys. 174 (1995)
57 and arXiv:hep-th/9405109;

G. Barnich, F. Brandt, M. Henneaux, Local BRST cohomology in the antifield formalism. II. Application to
Yang-Mills theory, Commun. Math. Phys. 174 (1995) 116 and
arXiv:hep-th/9405194;

G. Barnich, F. Brandt, M. Henneaux, General solution of the Wess-Zumino consistency condition for Einstein
gravity, Phys. Rev. D 51 (1995) R1435 and arXiv:hep-th/9409104;

S.D. Joglekar and B.W. Lee, General theory of renormalization of gauge
invariant operators, Ann. Phys. (NY) 97 (1976) 160.

\bibitem{lvgauge}  D. Anselmi and M. Halat, Renormalization of Lorentz
violating theories, Phys. Rev. D 76 (2007) 125011 and arXiv:0707.2480
[hep-th];

D. Anselmi, Weighted power counting and Lorentz violating gauge theories. I.
General properties, Ann. Phys. 324 (2009)\ 874, %
\href{http://renormalization.com/08a2/}{08A2 Renormalization.com} and
arXiv:0808.3470 [hep-th];

D. Anselmi, Weighted power counting and Lorentz violating gauge theories.
II. Classification, Ann. Phys. 324 (2009) 1058, %
\href{http://renormalization.com/08a3/}{08A3 Renormalization.com} and
arXiv:0808.3474 [hep-th].

\bibitem{weinberg}  S. Weinberg, Ultraviolet divergences in quantum theories
of gravitation, in \textit{An Einstein centenary survey}, Edited by S.
Hawking and W. Israel, Cambridge University Press, Cambridge 1979.

\bibitem{stelle}  K.S. Stelle, Renormalization of higher-derivative quantum
gravity, Phys. Rev. D 16 (1977) 953.

\bibitem{tombola}  E.T. Tomboulis, Superrenormalizable gauge and
gravitational theories, arXiv:hep-th/9702146;

L. Modesto, Super-renormalizable quantum gravity, Phys.Rev. D86 (2012)
044005 and arXiv:1107.2403 [hep-th];

T. Biswas, E. Gerwick, T. Koivisto and A. Mazumdar, Towards singularity and ghost free theories of gravity, Phys.Rev.Lett. 108 (2012) 031101 and arXiv:1110.5249 [gr-qc];

L. Modesto, Finite quantum gravity, arXiv:1305.6741 [hep-th].

\bibitem{kluberg}  H. Kluberg-Stern and J.B. Zuber, Renormalization of
nonabelian gauge theories in a background field gauge. 1. Green functions,
Phys. Rev. D12 (1975) 482;

H. Kluberg-Stern and J.B. Zuber, Renormalization of nonabelian gauge
theories in a background field gauge. 2. Gauge invariant operators, Phys.
Rev. D 12 (1975) 3159.

\bibitem{kostelecky}  D. Colladay and V.A. Kosteleck\'{y}, Lorentz-violating
extension of the Standard Model, Phys. Rev. D58 (1998) 116002 and
arXiv:hep-ph/9809521;

\bibitem{LVSM}  D. Anselmi, Weighted power counting, neutrino masses and
Lorentz violating extensions of the Standard Model, Phys. Rev. D79 (2009)
025017, \href{http://renormalization.com/08a4/}{08A4 Renormalization.com}
and arXiv:0808.3475 [hep-ph];

D. Anselmi, Standard Model without elementary scalars and high energy
Lorentz violation, Eur. Phys. J. C65 (2010) 523, %
\href{http://renormalization.com/09a1/}{09A1 Renormalization.com}, and
arXiv:0904.1849 [hep-ph].

\bibitem{adlerbardeen}  S.L. Adler and W.A. Bardeen, Absence of higher-order
corrections in the anomalous axial vector divergence, Phys. Rev. 182 (1969)
1517.

\bibitem{quadri}  D. Binosi and A. Quadri, Slavnov-Taylor constraints for
nontrivial backgrounds, Phys. Rev. D84 (2011) 065017 and arXiv:1106.3240
[hep-th];

D. Binosi and A. Quadri, Canonical transformations and renormalization group
invariance in the presence of nontrivial backgrounds, Phys. Rev. D85 (2012)
085020 and arXiv:1201.1807 [hep-th];

D. Binosi and A. Quadri, The background field method as a canonical
transformation, Phys.Rev. D85 (2012) 121702 and arXiv:1203.6637 [hep-th].

\bibitem{fieldcov}  D. Anselmi, A general field covariant formulation of
quantum field theory, Eur. Phys. J. C73 (2013) 2338, %
\href{http://renormalization.com/12a1/}{12A1 Renormalization.com} and
arXiv:1205.3279 [hep-th].

\bibitem{masterf} D. Anselmi, A master functional for quantum field theory, Eur. Phys. J. C73
(2013) 2385, \href{http://renormalization.com/12a2/}{12A2 Renormalization.com%
} and arXiv:1205.3584 [hep-th].

\bibitem{mastercan} D. Anselmi, Master functional and proper formalism for quantum gauge field
theory, Eur. Phys. J. C73 (2013) 2363, %
\href{http://renormalization.com/12a3/}{12A3 Renormalization.com} and
arXiv:1205.3862 [hep-th].

\bibitem{lavrov}  B.L. Voronov, P.M. Lavrov and I.V. Tyutin, Canonical
transformations and the gauge dependence in general gauge theories, Sov. J.
Nucl. Phys. 36 (1982) 292 and Yad. Fiz. 36 (1982) 498.

\bibitem{superg}  P. van Nieuwenhuizen, \textit{Supergravity}, Phys. Rept.
68 (1981) 189.

\bibitem{superspace}  S.J. Gates, M.T. Grisaru, M. Rocek and W. Siegel,
{\it Superspace or one thousand and one lessons in supersymmetry}, Front.Phys. 58
(1983) 1-548, arXiv:hep-th/0108200.

\bibitem{webbook}  D. Anselmi, \textit{Renormalization}, to appear at %
\href{http://renormalization.com}{\texttt{renormalization.com}}

\bibitem{BV2}  D. Anselmi, More on the subtraction algorithm, Class. Quant.
Grav. 12 (1995) 319, \href{http://renormalization.com/94a1/}{94A1
Renormalization.com} and arXiv:hep-th/9407023.

\bibitem{removal}  D. Anselmi, Removal of divergences with the
Batalin-Vilkovisky formalism, Class. Quant. Grav. 11 (1994) 2181-2204, %
\href{http://renormalization.com/93a2/}{93A2 Renormalization.com} and
arXiv:hep-th/9309085.

\bibitem{vanproeyen}  W. Troost, P. van Nieuwenhuizen and A. Van Proeyen,
Anomalies and the Batalin-Vilkovisky Lagrangian formalism, Nucl. Phys. B 333
(1990) 727.
\end{thebibliography}
\end{document}